\documentclass[aoas,MSNbibl,nameyear,dvips]{arximspdf}
\usepackage{dcolumn}
\usepackage{graphicx}
\usepackage{url,breakurl}

\doi{10.1214/14-AOAS759} 
\volume{8}
\issue{3}
\pubyear{2014}
\firstpage{1825}
\lastpage{1852}
\docsubty{FLA}

\makeatletter
\newcolumntype{d}[1]{D{.}{.}{#1}}
\newcommand{\llvert}{\vert}
\makeatother

\begin{document}
\begin{frontmatter}

\title{Nonstationary ETAS models for nonstandard~earthquakes}
\runtitle{Nonstationary ETAS models}

\begin{aug}
\author[A]{\fnms{Takao}~\snm{Kumazawa}\corref{}\thanksref{T1,m1}\ead[label=e1]{tkuma@ism.ac.jp}}
\and
\author[B]{\fnms{Yosihiko}~\snm{Ogata}\thanksref{T1,T3,m1,m2}}
\runauthor{T. Kumazawa and Y. Ogata}
\affiliation{The Institute of Statistical Mathematics\thanksmark{m1}
and University of Tokyo\thanksmark{m2}}
\address[A]{The Institute of Statistical Mathematics\\
10-3 Midori-cho, Tachikawa\\
Tokyo 190-8562\\
Japan\\
\printead{e1}} 
\address[B]{The Institute of Statistical Mathematics\\
10-3 Midori-cho, Tachikawa\\
Tokyo 190-8562\\
Japan\\
and\\
Earthquake Research Institute\\
University of Tokyo\\
1-1-1 Yayoi, Bunkyo-ku\\
Tokyo 113-0032\\
Japan}
\end{aug}
%
\thankstext{T1}{Supported by JSPS KAKENHI Grant Numbers 23240039 and 26240004,
supported by a postdoctoral fellowship from the Institute of Statistical Mathematics.}
\thankstext{T3}{Supported by the Aihara Innovative Mathematical
Modelling Project,
the ``Funding Program for World-Leading Innovative R\&D on Science and
Technology (FIRST Program),'' initiated by the Council for Science and
Technology Policy.}

\received{\smonth{10} \syear{2013}}
\revised{\smonth{4} \syear{2014}}

%
\begin{abstract}
The conditional intensity function of a point process is a useful tool for
generating probability forecasts of earthquakes. The epidemic-type
\mbox{aftershock} sequence (ETAS) model is defined by a conditional intensity
function, and the corresponding point process is equivalent to a branching
process, assuming that an earthquake generates a cluster of offspring
earthquakes (triggered earthquakes or so-called aftershocks). Further, the
size of the first-generation cluster depends on the magnitude of the
triggering (parent) earthquake. The ETAS model provides a good fit to
standard earthquake occurrences. However, there are nonstandard earthquake
series that appear under transient stress changes caused by aseismic forces
such as volcanic magma or fluid intrusions. These events trigger transient
nonstandard earthquake swarms, and they are poorly fitted by the stationary
ETAS model. In this study, we examine nonstationary extensions of the ETAS
model that cover nonstandard cases. These models allow the parameters
to be
time-dependent and can be estimated by the empirical Bayes method. The best
model is selected among the competing models to provide the inversion
solutions of nonstationary changes. To address issues of the uniqueness and
robustness of the inversion procedure, this method is demonstrated on an
inland swarm activity induced by the 2011 Tohoku-Oki, Japan earthquake of
magnitude~9.0.
\end{abstract}

%
\end{frontmatter}

\section{Introduction}\label{sec1}
The epidemic-type aftershock sequence (ETAS) model
[\citeauthor{OGA85} (\citeyear{OGA85,OGA86,OGA88,OGA89})] is one of the
earliest point-process
models created for clustered events. It is defined in terms of a
conditional intensity [\citet{Haw71}, \citet{HAWADA73},
\citeauthor{Oga78}
(\citeyear{Oga78,OGA81})], is equivalent to epidemic branching
processes [\citet{Ken49}, \citet{HawOak74}], and allows each
earthquake to generate
(or trigger) offspring earthquakes. Besides being used in seismology, the
ETAS model has been applied to various fields in the social and natural
sciences [e.g., \citet{Baletal12}, \citet{CHAMCG12},
\citet{HAS}, \citet{HERSCH09}, \citet{Mohetal11}, \citet{PenSchWoo05}, \citet{SCHPENWOO03}].

Similar magnitude-dependent point-process models have been applied to
seismological studies [\citet{VERDAV66}, \citet{LOM74}, \citet{KagKno87}] and statistical studies [\citet{Ver70}]. The ETAS
model is stationary if the immigration rate (background seismicity
rate) of
an earthquake remains constant and the branching ratio is subcritical
[\citet{Haw71}, \citet{HawOak74}, \citet{ZHUOGA06}].

The history-dependent form of the ETAS model on occurrence times and sizes
(magnitudes) lends itself to the accumulated empirical studies by
\citeauthor{UTS61} (\citeyear{UTS61,UTS62,UTS69,UTS70,UTS71,UTS72}) and
\citet{UTSSEK55}, and its establishing
history is detailed by \citet{UTSOGAMAT95}. ETAS model parameters can be
estimated from earthquake occurrence data by maximizing the log-likelihood
function to provide estimates for predicting seismic activity (i.e., number
of earthquakes per unit time). The model has been frequently used and cited
in seismological studies, especially to compare the features of simulated
seismicity with those of real seismicity data. The model is also
recommended for use in short-term predictions [\citet{JORCHEGAS12}] in the
report of the International Commission on Earthquake Forecasting for Civil
Protection. It is planned to be adopted for operational forecasts of
earthquakes in \mbox{California} (The Uniform California Earthquake Rupture
Forecast, Version 3, URL:
\url{http://www.wgcep.org/sites/wgcep.org/files/UCERF3\_Project\_Plan\_v55.pdf}).

The ETAS model has also been used to detect anomalies such as
quiescence in
seismicity. Methods and applications are detailed in
\citeauthor{OGA88} (\citeyear
{OGA88,OGA89,OGA92,OGA99,OGA,OGA06N1,OGA07,OGA10,OGA11N1,OGA12}),
\citet{OGAJONTOD03},
\citet{KUMOGATOD10}, and \citet{BANOGA13}. A change-point
analysis examines a simple hypothesis that specific parameters change after
a certain time. The misfit of occurrence rate prediction after a change
point is then \mbox{preliminarily} shown by the deviation of the empirical
cumulative counts of the earthquake occurrences from the predicted
cumulative function. The predicted function is the extrapolation of the
model fitted before the change point. A~downward and upward deviation
corresponds to relative quiescence and activation, respectively.

This study considers a number of nonstationary extensions of the ETAS model
to examine more detailed nonstandard transient features of earthquake
series. The extended models take various forms for comparison with the
reference ETAS model, which represents the preceding normal activity in a
given focal region. Because changing stresses in the crust are not directly
observable, it is necessary to infer relevant quantitative characteristics
from seismic activity data. For example, \citet{HAIOGA05} and
\citet{LOMCOCMAR10} estimated time-dependent background rates
(immigration rates) in a moving time window by removing the triggering
effect in the ETAS model.

In Section~\ref{sec2}, time-dependent parameters for both background rates and
productive rates are simultaneously estimated. There, the penalized
log-likelihood is considered for the trade-off between a better fit of the
nonstationary models and the roughness penalties against overfitting. Then,
not only is an optimal strength adjusted for each penalty but also a better
penalty function form is selected using the Akaike Bayesian Information
Criterion (\textit{ABIC}) [\citet{Aka80}]. These parameter constraints
together with
the existence of a change point are further examined to determine if they
improve the model fit. One benefit of this model is that it allows varying
parameters to have sharp changes or discontinuous jumps at the change point
while sustaining the smoothness constraints in the rest of the period.

In Section~\ref{sec3}, the methods are demonstrated by applying the model to a swarm
activity. The target activity started after the March 11, 2011 Tohoku-Oki
earthquake of magnitude (M) 9.0, induced at a distance from the M9.0
rupture source. Section~\ref{sec4} concludes and discusses the models and methods.
The reproducibility of the inversion results is demonstrated in the
\hyperref[sec6]{Appendix} by synthesizing the data and re-estimating it using the same
procedure.

\section{Methods}\label{sec2}
\subsection{The ETAS model}\label{sec21} A conditional intensity
function characterizes
a point (or counting) process $N(t)$ [\citet{DALVER03}]. The
conditional intensity $\lambda(t|H_t)$ is defined as follows:
%
\begin{equation}
\label{equ1} \Pr \bigl\{ N(t,t + dt) = 1|H_{t} \bigr\} = \lambda
(t|H_{t})\,dt + o(dt),
\end{equation}
where $H_{t}$ represents the history of occurrence times of marked events
up to time~$t$. The conditional intensity function is useful for the
probability forecasting of earthquakes, which is obtained by integrating
over a time interval.

The ETAS model, developed by \citeauthor{OGA85} (\citeyear
{OGA85,OGA86,OGA88,OGA89}), is a special
case of the marked Hawkes-type self-exciting process, and has the following
specific expression for conditional intensity:
%
\begin{equation}
\label{equ2} \lambda_{\theta} (t|H_{t}) = \mu+ \sum
_{ \{ i\dvtx S < t_{i} < t
\}} \frac{K_{0}e^{\alpha( M_{i} - M_{z} )}}{ ( t -
t_{i} + c )^{p}},
\end{equation}
where $S$ is the starting time of earthquake observation and $M_{z}$
represents the smallest magnitude (threshold magnitude) of earthquakes to
be treated in the data set. $M_{i}$ and $t_{i}$ represent the magnitude and
the occurrence time of the $i$th earthquake, respectively, and $H_{t}$
represents the occurrence series of the set ($t_{i}$, $M_{i}$) before time~$t$. The parameter set $\theta$ thus consists of five elements ($\mu$,
$K_{0}, c, \alpha, p$). In fact, the second term of equation~(\ref{equ2}) is a
weighted superposition of the Omori--Utsu empirical function
[\citet{UTS61}]
for aftershock decay rates,
%
\begin{equation}
\label{equ3} \lambda_{\theta} (t) = \frac{K}{ ( t + c )^{p}},
\end{equation}
where $t$ is the elapsed time since the main shock. It is important to note
that, while the concept of a main shock and its aftershocks is intuitively
classified by seismologists sometime after the largest earthquake occurs,
there is no clear discrimination between them in equation (\ref{equ2}). That is,
each earthquake can trigger aftershocks, and the expected cluster size
depends on the magnitude of the triggering earthquake with the parameter
$\alpha$.

The parameter $K_{0}$ (earthquakes/day) is sometimes called the
``aftershock productivity.'' As the name explains, the parameter controls
the overall triggering intensity. The factor $c$ (day) is a scaling
time to
establish the power-law decay rate and allows a finite number of
aftershocks at the origin time of a triggering earthquake (a main shock).
In practice, the fitted values for $c$ are more likely to be caused by the
under-reporting of small earthquakes hidden in the overlapping wave trains
of large earthquakes [\citet{UTSOGAMAT95}]. The exponent $p$ is the
power-law decay rate of the earthquake rate in equation (\ref{equ3}). The magnitude
sensitivity parameter $\alpha$ (magnitude $^{-1}$) accounts for the
efficiency of an earthquake of a given magnitude in generating aftershocks.
A small $\alpha$ value allows a small earthquake to trigger a larger
earthquake more often. Finally, the background (spontaneous) seismicity
rate $\mu$ represents sustaining external effects and superposed occurrence
rates of long-range decays from unobserved past large earthquakes. It also
accounts for the triggering effects by external earthquakes.

The FORTRAN program package associated with manuals regarding\break ETAS analysis
is available to calculate the maximum likelihood estimates (MLEs)
of~$\theta$ and to visualize model performances [\citet{OGA06N2}]. See
also \url{http://www.ism.ac.jp/\textasciitilde ogata/Ssg/ssg\_softwaresE.html}.

\subsection{Theoretical cumulative intensity function and time transformation}\label{sec22} Suppose that the parameter values $\theta
= (\mu, K, c,
\alpha, p)$ of the ETAS, equation (\ref{equ2}), are given. The integral of the
conditional intensity function,
%
\begin{equation}
\label{equ4} \Lambda_{\theta} (t|H_{t}) = \int
_{S}^{t} \lambda_{\theta}
(u|H_{u}) \,du,
\end{equation}
provides the expected cumulative number of earthquakes in the time interval
$[0, t]$. The time transformation from $t$ to $\tau$ is based on the
cumulative intensity,
%
\begin{equation}
\label{equ5} \tau= \Lambda(t|H_{t}),
\end{equation}
which transforms the original earthquake occurrence time $(t_{1}, t_{2},\ldots, t_{N})$ into the sequence $(\tau_{1}, \tau_{2},\ldots, \tau
_{N})$ in
the time interval $[0,\Lambda(T)]$. If the model represents a good
approximation of the real seismicity, it is expected that the integrated
function [equation~(\ref{equ4})] and the empirical cumulative counts $N(t)$ of the
observed earthquakes are similar. This implies that the transformed
sequence appears to be a \mbox{stationary} Poisson process (uniformly distributed
occurrence times) if the model is sufficiently correct, and appears to be
heterogeneous otherwise.

\subsection{Two-stage ETAS model and the change-point problem}\label
{sec23} In
change-point analysis, the whole period is divided into two disjointed
periods to fit the ETAS models separately, and is therefore called a
two-stage ETAS model. This is one of the easiest ways to treat
nonstationary data, and is best applied to cases in which parameters are
suspected to change at a specific time. Such a change point is observed
when a notably large earthquake or slow slip event (regardless of observed
or unobserved) occurs in or near a focal region. Many preceding studies
[e.g., \citet{OGAJONTOD03}, \citeauthor{OGA} (\citeyear
{OGA,OGA06N1,OGA07,OGA10}), \citet{KUMOGATOD10}] have adopted this
method to their case studies, and details can
be found therein.

The question of whether the seismicity changes at some time $T_{0}$ in a
given period $[S, T]$ is reduced to a problem of model selection. In this
analysis, the ETAS models are separately fitted to the divided periods $[S,
T_{0}]$ and $[T_0, T]$, and their total performance is compared
to an ETAS model fitted over the whole period $[S, T]$ by the Akaike Information Criterion (\textit{AIC}) [Akaike (\citeyear{Ak73,Ak74,Ak77})]. The
\textit{AIC} is described as follows:
%
\begin{equation}
\label{equ6} \mathit{AIC} = - 2\max\log L(\theta) + 2k,
\end{equation}
where ln $L(\theta)$ represents the log-likelihood of the ETAS model,
%
\begin{equation}
\label{equ7} \log L(\theta) = \sum_{ \{ i\dvtx S < t_{i} < T \}} \log
\lambda_{\theta} ( t_{i}| H_{t_{i}} ) - \int
_{S}^{T} \log\lambda_{\theta} ( t|
H_{t} )\,dt,
\end{equation}
and $k$ is the number of parameters to be estimated. The variables $t_{i}$
and $H_{t_i}$ are the same as those in equation (\ref{equ2}).
Under this criterion, the model with a smaller \textit{AIC} value performs
better. It is useful to keep in mind that $\exp\{- \Delta\mathit{AIC}/2\}$
can be interpreted as the relative probability of how a model with a
smaller \textit{AIC} value is superior to others [e.g., \citet{Aka80}].

Let \textit{AIC}$_{0}$ be the \textit{AIC} of the ETAS model estimated for
the whole period $[S, T]$, $\mathit{AIC}_{1}$ be that of the first period
$[S, T_{0}]$, and $\mathit{AIC}_{2}$ be that of the second period $[T_0,
T]$, therefore,
%
\begin{eqnarray}
\label{equ8} \mathit{AIC}_{0} &=& - 2\max_{\theta_{0}}\log L(
\theta_{0};S,T) + 2k_{0},\nonumber
\\
\mathit{AIC}_{1} &=& - 2\max_{\theta_{1}}\log L(
\theta_{1};S,T_{0}) + 2k_{1},
\\
\mathit{AIC}_{2} &=& - 2\max_{\theta_{2}}\log L(
\theta_{2};T_{0},T) + 2k_{2}.\nonumber
\end{eqnarray}

Let $\mathit{AIC}_{12}$ represent the total $\mathit{AIC}$ from the divided periods,
such that
%
\begin{equation}
\label{equ9} \mathit{AIC}_{12} = \mathit{AIC}_{1} + \mathit{AIC}_{2} + 2q,
\end{equation}
with $q$ being the degrees of freedom to search for the best change-point
candidate $T_{0}$. Next, $\mathit{AIC}_{12}$ is compared against
\textit{AIC}$_{0}$. If $\mathit{AIC}_{12}$ is smaller, the two-stage ETAS
model with the change point $T_{0}$ fits better than the ETAS model applied
to the whole interval. The quantity $q$ monotonically depends on sample
size (number of earthquakes in the whole period $[S, T])$ when searching
for the maximum likelihood estimate of the change point [\citeauthor
{OGA92}
(\citeyear{OGA92,OGA99}), \citet{KUMOGATOD10}, \citet
{BANOGA13}]. This penalty term
$q$, as well as an increased number of estimated parameters, imposes a
hurdle for a change point to be significant, and it is usually rejected
when the one-stage ETAS model fits sufficiently well. If the change point
$T_{0}$ is predetermined from some information other than the data,
then $q
= 0$. This is often the case when a conspicuously large earthquake occurs
within swarm activity, and will be discussed below. Also, even in this
case, the overfitting by the change point is avoided by the $\mathit{AIC}_{12}$
of a
two-stage ETAS model, which has two times as many parameters of a single
stationary ETAS model throughout the whole period.

\subsection{Anomaly factor functions for nonstationary ETAS models}\label{sec24}
\begin{figure}

\includegraphics{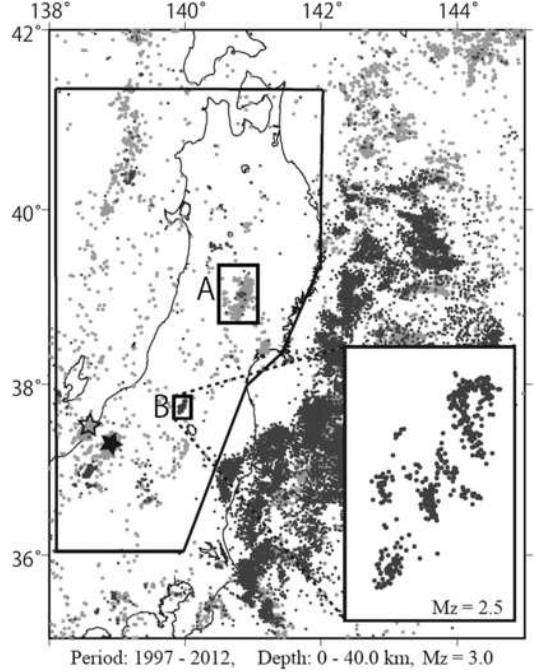}

\caption{Epicenters of earthquakes of magnitude (M)${}\geq 3.0$ in
the Northern Honshu region, Japan, with depths shallower than 40 km, from
1997 to 2012, selected from the JMA Hypocenter catalog. The gray and black
dots represent the earthquakes that occurred before and after the M9.0
Tohoku-Oki earthquake, respectively. The rectangular regions A and B
include the aftershocks of the 2008 Iwate-Miyagi Prefectures Inland
Earthquake of M7.2 and the swarm near Lake Inawashiro, respectively. Their
inset panels magnify the epicenter distribution with M${}\geq 2.0$
and M${}\geq 2.5$, respectively. The polygonal region indicates the
Tohoku inland and its western offshore region; the earthquakes in this
region are used in the reference stationary ETAS model. The closed star
represents the epicenter of the 2004 Chuetsu earthquake of M6.8, and the
open star represents the 2007 Chuetsu-Oki earthquake of M6.8.}\label{fig1}
\end{figure}

\begin{figure}

\includegraphics{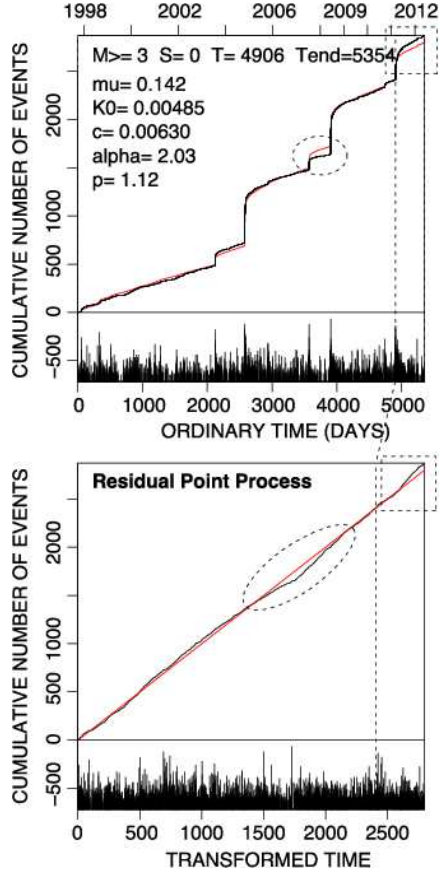}

\caption{Cumulative number and magnitude of earthquakes of
M${}\geq 3$ against the ordinary time and transformed time by the
ETAS model from the polygonal region in Figure~\protect\ref{fig1}. The fitted period of the
model is from October 1997 to the M9.0 March 2011 Tohoku-Oki earthquake
(indicated by vertical dashed lines). Red curves in the top and bottom
panels represent the theoretical cumulative numbers against the ordinary
time (\protect\ref{equ4}) and the transformed time, respectively. The dashed black ellipses
and dashed rectangles highlight the anomalies around 2008 and after the
Tohoku-Oki earthquake, respectively.}\vspace*{-2pt}\label{fig2}
\end{figure}

Assume
that the ETAS model fits the data well for a period of ordinary seismic
activity. Then, the concern is whether this model shows a good fit to the
seismicity in a forward extended period. If there are misfits,
time-dependent compensating factors are introduced to the parameters to be
made time-dependent. These factors are termed ``anomaly factor functions''
and, thus, the transient changes in parameters are tracked. If earthquake
activity is very low in and near a target region preceding the transient
activity, data from a wider region, such as the polygonal region in
Figure~\ref{fig1}, is used to obtain a reference stationary ETAS model
(Figure~\ref{fig2}). Such a
model is stable against small local anomalies, and is therefore a good
reference model. The reference ETAS model, coupled with the corresponding
anomaly factor functions, becomes the nonstationary ETAS model in this
study.

%
%
%
%

Among the parameters of the ETAS model, the background rate $\mu$ and the
aftershock productivity $K_{0}$ are sensitive to nonstationarity. We
therefore introduce the anomaly factor functions as the nonstationary
components to modify the reference stationary ETAS model in such a way
that
%
\begin{equation}
\label{equ10} \lambda_{\theta} (t|H_{t}) = \mu q_{\mu}
(t) + \sum_{\{ i\dvtx  S < t_{i} <
t\}} \frac{K_{0}q_{K}(t_{i})e^{\alpha( M_{i} - M_{z} )}}{ (
t - t_{i} + c )^{p}}.
\end{equation}

Here $k_\mu(t)$ and $q_K (t)$ are
referred to as anomaly factor functions of the parameters $\mu$ and
$K_{0}$, respectively. Because of technical reasons to avoid further model
complexity, we did not consider the case in which the other three
parameters $c$, $\alpha$ and $p$ in equation (\ref{equ2}) also are time-varying.
One structural problem of the ETAS model is that $K_{0}$ is correlated with
the parameter $\alpha$. The trade-off is not negligible, especially
when the
range of magnitudes in the data set is small. See Section~\ref{sec4} for additional
discussion of this issue.

We use the first-order spline function of the ordinary time $t$. This
is a
broken line interpolated by the coordinates $\{(t_{i}, q_{i}); i = 0,
1, 2,\ldots, N+1\}$, where $t_{i}$ is the occurrence time of the $i$th
earthquake, and $t_{0}$ and $t_{N+1}$ are the start and end of the period,
respectively. Then, the spline functions are defined as follows:
%
\begin{equation}
\label{equ11} \quad q_{\mu} (t) = \sum_{i = 1}^{N}
I_{(t_{i},t_{i + 1})}(t) \biggl\{ \frac{q_{\mu,i + 1} - q_{\mu,
i}}{t_{i + 1} - t_{i}}(t - t_{i}) +
q_{\mu,
i} \biggr\} = \sum_{i = 1}^{N}
q_{\mu, i}F_{i} (t)
\end{equation}
and
%
\begin{equation}
\label{equ12} \quad q_{K}(t) = \sum_{i = 1}^{N}
I_{(t_{i},t_{i + 1})}(t) \biggl\{ \frac{q_{K, i + 1} - q_{K, i}}{t_{i +
1} - t_{i}}(t - t_{i}) +
q_{K, i} \biggr\} = \sum_{i = 1}^{N}
q_{K, i}F_{i} (t),
\end{equation}
where $I_{(t_{i},t_{i + 1})}(t)$ is the indicator function, with the explicit
form of $F_{i} (t)$ given as
%
\begin{equation}
\label{equ13} F_{i}(t) = \frac{t - t_{i - 1}}{t_{i} - t_{i - 1}}I_{(t_{i -
1},t_{i})}(t) +
\frac{t_{i + 1} - t}{t_{i + 1} - t_{i}}I_{(t_{i},t_{i +
1})}(t).
\end{equation}
The log-likelihood function of the nonstationary point process can be
written as follows:
%
\begin{equation}
\label{equ14} \log L(q) = \sum_{\{ i; S < t_{i} < T\}} \log
\lambda_{q}(t_{i}|H_{t_{i}}) - \int
_{S}^{T} \lambda_{q} (t|H_{t})
\,dt,
\end{equation}
where $q = (q_{\mu}, q_{K})$.

\subsection{Penalties against rough anomaly factor functions}\label{sec25}
Since these
ano\-maly functions have many coefficients representing flexible variations,
coefficients are estimated under an imposed smoothness constraint to avoid
their overfitting. This study uses the penalized log-likelihood
[\citet{GooGas71}] described below. With the roughness penalty functions,
%
\begin{eqnarray}
\label{equ15} \Phi_{\mu} &=& \sum_{i = 0}^{N}
\biggl( \frac{q_{\mu,i + 1} - q_{\mu,
i}}{t_{i + 1} - t_{i}} \biggr)^{2}(t_{i + 1} -
t_{i})\quad\mbox{and}
\nonumber
\\[-8pt]
\\[-8pt]
\Phi_{K} &=& \sum_{i = 0}^{N}
\biggl( \frac{q_{K,i + 1} - q_{K, i}}{t_{i + 1}
- t_{i}} \biggr)^{2}(t_{i + 1} -
t_{i}),
\nonumber
\end{eqnarray}
and the penalized log-likelihood against the roughness becomes
%
\begin{equation}
\label{equ16} Q ( q| w_{\mu},w_{K} ) = \log L ( q ) -
w_{\mu} \Phi_{\mu} - w_{K}\Phi_{K},
\end{equation}
where each ``$w$'' represents weight parameters that tune the smoothness
constraints of the anomaly factors. The roughness penalty, equation (\ref{equ15}),
imposes penalties to the log-likelihood according to parameter
differentials at successive event occurrence times.

Furthermore, the degree of the smoothness constraints may not be
homogeneous in ordinary time because earthquake series are often highly
clustered. In other words, it is expected that more detailed or rapid
changes of the anomaly factors appear during dense event periods rather
than during sparse periods [\citet{OGA89}, \citet{AdeOga10}].
Hence, for the same model, alternative constraints are considered by
replacing $\{t_{i}\}$ in equation (\ref{equ15}) with $\{ \tau_{i}\}$ on the
transformed time $\tau$ in equation (\ref{equ5}) of the reference ETAS model.

The following restricted cases of the nonstationary model in equation (\ref{equ10}),
together with different types of the aforementioned parameter constraints,
are examined and summarized in Table~\ref{tab1}. Model~1 restricts the parameter
$K_{0}$ to be constant and unchanged from the reference model, leaving
$q_{\mu} (t)$ to be unrestricted. Model~2 restricts the parameters $\mu$
and $K_{0}$ to have the same factor. In other words, model~2 estimates the
anomaly factor for the total intensity $\lambda_{\theta} (t|H_{t})$ in
equation (\ref{equ10}). This restriction is assumed in \citet{AdeOga10}.
Model~3 has no restriction.

%
\begin{table}[t]
\tabcolsep=0pt
\caption{Summary of the competing nonstationary ETAS models. The
numbers index the models. The row headers explain the model
restrictions of
anomaly factors $q_{\mu}(t)$ and $q_{K}(t)$. The
first column \textup{(a)} uses smoothing on ordinary time, the second column \textup{(b)} on
the transformed time}\label{tab1}
\begin{tabular*}{\tablewidth}{@{\extracolsep{\fill}}@{}lcc@{}}
\hline
\textbf{Restrictions} & \textbf{(a) Smoothing on ordinary time} & \textbf{(b) Smoothing on transformed time}\\
\hline
$q_{K} (t) = 1$ & Model 1(a) & Model 1(b)\\
$q_{\mu} (t) = q_{K}(t)$ & Model 2(a) & Model 2(b)\\
No restriction & Model 3(a) & Model 3(b)\\
\hline
\end{tabular*}
\end{table}

Here, from a statistical modeling viewpoint, it should be noted that
$\mu$
and $K_{0}$ are linearly parameterized regarding the conditional intensity
[equation~(\ref{equ2})], and likewise the linearly parameterized coefficients of the
functions $q_{\mu}$ and $q_{K}$ in equation (\ref{equ10}). Together, they force the
penalized log-likelihood function [equation~(\ref{equ16})] to be strictly concave
regardless of the dimensions of the coefficients' space [\citeauthor
{Oga78} (\citeyear{Oga78,Oga01}),
\citet{OGAKAT93}]. Therefore, the maximizing solutions of the
penalized log-likelihood function can be obtained uniquely and stably under
a suitable numerical optimization algorithm [e.g., appendices of
\citeauthor{OGA04} (\citeyear{OGA04,OGA11N2})]. The reproducibility of
the inversion results of $\mu(t)$ and
$K_{0} (t)$ are demonstrated in the \hyperref[sec6]{Appendix}.

\subsection{Tuning smoothness constraints, model selection and error evaluation}\label{sec26}

In a Bayesian context, given the weights, the solution of the parameters
$q$ that minimize the penalized log-likelihood $Q$ in (\ref{equ16}) is termed the
maximum a posteriori (MAP) estimate. In the following section, we describe
how to determine the optimal MAP (OMAP) estimate. To obtain the optimal
weights in the penalty \mbox{functions} in equation (\ref{equ16}), this study uses a
Bayesian interpretation of penalized log-likelihood as suggested by
\citet{Aka80}. Specifically, the exponential of each penalty
function is
proportional to a prior Gaussian distribution of the forms
%
\begin{equation}
\label{equ17} \pi( q_{\mu} | w_{\mu} ) \propto e^{ -
w_{\mu} q_{\mu} \Sigma_{\mu} q_{\mu}^{t}/2}
\quad \mbox{and}\quad\pi( q_{K}| w_{K} ) \propto
e^{ -
w_{K}q_{K}\Sigma_{K}q_{K}^{t}/2},
\end{equation}
since the coefficients of the function $q_{.}(\cdot)$ in the penalty term
$\Phi$ take a quadratic form with a symmetric $(N+1) \times(N+1)$
nonnegative definite matrix $\Sigma$. Since each matrix $\Sigma$ is
degenerate and has $\operatorname{rank} (\Sigma) = N$, above each prior
distribution becomes improper [\citet{OGAKAT93}]. To avoid such improper
priors, we divide each of the vectors $q$ into $(q^{c}, q^{(N+1)})$ so that
each of the priors becomes a probability density function with respect to
$q^{c}$:
%
\begin{equation}
\label{equ18} \pi \bigl( q^{c}| w, q_{N + 1} \bigr) =
\frac{ (
w^{N}\det\Sigma^{c} )^{1 / 2}}{\sqrt{2\pi}^{N}}\exp \biggl( - \frac{1}{2}w^{N}q^{c}
\Sigma^{c} {}^{t}q^{c} \biggr),
\end{equation}
where $\Sigma^{c}$ is the cofactor of the last diagonal element of
$\Sigma$, and $w$ and $q^{(N+1)}$ are considered hyperparameters to
maximize the integral of the posterior distribution with respect to
$q^{c}$,
%
%
\begin{eqnarray}\label{equ19}
&& \Psi \bigl( w_{\mu},w_{K};
q_{\mu}^{(N + 1)}, q_{K}^{(N + 1)} \bigr)
\nonumber\\[-8pt]\\[-8pt]\nonumber
&&\qquad = \int L
( q_{\mu}, q_{K} ) \pi( q_{\mu} | w_{\mu} )
\pi( q_{K}| w _{K} )\,dq_{\mu}^{c}\,dq_{K}^{c},
\end{eqnarray}
which refers to the likelihood of a Bayesian model. \citet{Goo65} suggests
the maximization of equation (\ref{equ19}) with respect to the hyperparameters and
termed this the Type II maximum likelihood procedure.

By applying Laplace's method [\citet{Sti86}, pages 366--367], the posterior
distribution is approximated by a Gaussian distribution, by which the
integral in equation (\ref{equ19}) becomes
%
\begin{eqnarray}\label{equ20}
&& \Psi \bigl( w_{\mu},w_{K};
q_{\mu}^{(N + 1)}, q_{K}^{(N + 1)} \bigr)\nonumber
\\
&&\qquad = Q
\bigl( \hat{q}_{\mu}^{c}, \hat{q}_{K}^{c}
| w_{\mu}, w_{K}; q_{\mu}^{(N + 1)},
q_{K}^{(N + 1)} \bigr)
\\
&&\quad\qquad{} - \tfrac{1}{2}\log( \det H_{\mu} ) - \tfrac{1}{2}\log( \det
H_{K} ) + MN\log2\pi,\nonumber
\end{eqnarray}
where $\hat{q}$ is the maximum of the penalized log-likelihood $Q$ in
equation (\ref{equ16}) and
%
\begin{equation}
\label{equ21} H \bigl( \hat{q}{}^{c}| w, q^{(N + 1)} \bigr) =
\frac{\partial^{2}\log L ( \hat{q}^{c}\llvert w, q^{(N + 1)}
)}{\partial q^{c}\,\partial(q^{c})^{t}} - \Sigma^{c} \bigl(w, q^{(N +
1)} \bigr),
\end{equation}
for a fixed weight $w$ for either $w_{\mu}$ or $w_{K}$.

Thus,\vspace*{1pt} maximizing equation (\ref{equ16}) with respect to $q^{c}$ and equation (\ref{equ20})
with respect to $ ( w_{\mu},w_{K}; q_{\mu}^{(N + 1)}, q_{K}^{(N + 1)}
)$, in turn, achieves our objective. In the former maximization, a
quasi-Newton method using the gradients $\partial\log L ( q ) /
\partial q$ and the Newton method making use of the Hessian matrices,
equation (\ref{equ21}), endure a fast convergence regardless of high dimensions. For
the latter maximization, a direct search such as the simplex method is used.
A flowchart of numerical algorithms is described in the appendices of
\citeauthor{OGA04}
(\citeyear{OGA04,OGA11N2}).

Anomaly factor functions under the optimal roughness penalty result in
suitably smooth curves throughout the period. Furthermore, there may be a
change point that results in sudden changes in parameters $\mu$ or $K$. To
examine such a discontinuity, a sufficiently small weight is put into the
interval that includes a change point (e.g., $w = 10^{-5}$), and the
goodness-of-fit by \textit{ABIC} is compared with that of the smooth model
with the optimal weights for all intervals.

It is useful to obtain the estimation error bounds of the MAP estimate
$\hat q$ at each time of an observed earthquake. The
joint error distribution of the parameters
at $\hat q$ is nearly a $2N$-dimensional\vspace*{1pt} normal
distribution $N(0, H^{- 1})$, where $H^{- 1} = (h^{ i, j})$, and $H =
(h_{ i, j})$ is the Hessian matrix in equation (\ref{equ21}). Hence, the covariance
function of the error process becomes
%
\begin{equation}
\label{equ22} c ( u,v ) = \sum_{i = 1}^{2N}
\sum_{j = 1}^{2N} F_{i} ( u
)h^{i,j}F_{j} ( v ),
\end{equation}
where $F_{i} = F_{ N+ i}$ for $i = 1, 2,\ldots, N$, which is defined in
equation (\ref{equ13}). Thus, the standard error of $q$ is provided by
%
\begin{equation}
\label{equ23} \varepsilon( t ) = \bigl[ \varepsilon_{\mu} ( t ),
\varepsilon_{K} ( t ) \bigr] = \sqrt{C ( t,t )}.
\end{equation}

\subsection{Bayesian model comparison}\label{sec27} It is necessary to
compare the
goodness of fit among the competing models. From equation (\ref{equ20}), the \textit{ABIC}
[\citet{Aka80}] can be obtained as
%
\begin{eqnarray}
\label{equ24} \mathit{ABIC} &=& ( - 2 )\max_{w_{\mu},w_{K}; q_{\mu}^{(N + 1)},
q_{K}^{(N + 1)}}\log\Psi \bigl(
w_{\mu},w_{K}; q_{\mu}^{(N + 1)},
q_{K}^{(N + 1)} \bigr)
\nonumber
\\[-8pt]
\\[-8pt]
&&{} + 2 \times( \# \mathrm{hyperparameter} ).
\nonumber
\end{eqnarray}
Specifically, models 1 and 2 [1(a)~and~(b), 2(a)~and~(b) in Table~\ref{tab1}] have four
hyperparameters, and model 3 [3(a)~and~(b)] has eight. A Bayesian model with
the smallest \textit{ABIC} value provides the best fit to the data.

Since there are various constraints in the different setups, the resulting
\textit{ABIC} values cannot be simply compared because of unknown different
constants, mainly due to the approximations in equation (\ref{equ20}).
Alternatively, the difference of \textit{ABIC} values relative to those
corresponding to the reference model are used. In other words, the
reduction amount of the \textit{ABIC} value from a very heavily constrained
case,
%
\begin{equation}
\label{equ25} \Delta \mathit{ABIC} = \mathit{ABIC} - \mathit{ABIC}_{0},
\end{equation}
where \textit{ABIC} is that of equation (\ref{equ24}) and $\mathit{ABIC}_{0}$ is
the \textit{ABIC} value with very heavy fixed weights, which constrain the
function to be almost constant. Therefore, the $\Delta\mathit{ABIC}$
approximates the \textit{ABIC} improvement from the flat anomaly functions
[$q(t) = 1$ for all $t$] to the optimal functions.

Likewise in \textit{AIC}, it is useful to keep in mind that exp$\{-
\Delta
\mathit{ABIC}/2\}$ can be interpreted as the relative probability of how
the model with the smallest \textit{ABIC} value is superior to others
[e.g., \citet{Aka80}].

\section{Applications}\label{sec3}
\subsection{The stationary ETAS model versus the two-stage ETAS model}\label{sec31}
First, we estimate the stationary ETAS model that has been applied to a
series of earthquakes of magnitude (M) 3.0 and larger contained in the
polygonal region highlighted in Figure~\ref{fig1}, from October 1997 to the M9.0
Tohoku-Oki earthquake on March 11, 2011. Specifically, the MLE has been
obtained for the stationary ETAS model [equation~(\ref{equ2})] by applying a normal
activity for earthquakes of M3.0 and larger from October 1997 to March 10,
2011 (Figure~\ref{fig2}). According to the estimated theoretical cumulative
curve in
ordinary time [equation~(\ref{equ4})] and transformed time [equation(\ref{equ5})] in Figure~\ref{fig2},
the ETAS model appears to fit very well except for a period near 2008
and a
period after the Tohoku-Oki earthquake, which is in good accordance with
\citet{OGA12}. These anomalies are highlighted by dashed ellipses
and dashed
rectangles in Figure~\ref{fig2}.

The former is the apparent lowering due to substantially small productivity
in the aftershock activity of the 2007 Chuetsu-Oki earthquake (open
star in
Figure~\ref{fig1}). Interestingly enough, the 2004 Chuetsu earthquake (closed star)
and the 2007 Chuetsu-Oki earthquake, which are about 40 km apart, have the
same magnitude (M6.8), but the number of aftershocks of $M\geq4.0$
differs by 6--7 times [\citet{Age}].

The latter is due to the activation relative to the predicted ETAS model.
The March 11, 2011 M9.0 Tohoku-Oki earthquake induces this activation. On
the other hand, a series of aftershocks (located in region~A, Figure~\ref{fig1}) of
the 2008 M7.2 Iwate-Miyagi Prefecture inland earthquake is quiet relative
to the occurrence rate predicted by the ETAS model estimated from the
aftershock data before the M9.0 earthquake.

%
%
\begin{figure}

\includegraphics{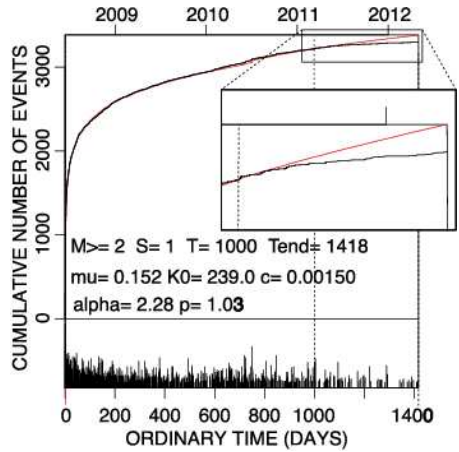}

\caption{Cumulative number and magnitude of the aftershock sequence
with $M\geq 1.5$, following the 2008 Iwate-Miyagi earthquake of
M7.2, from the region A against ordinary time. The ETAS model is fitted to
the sequence for the period from one day after the main shock ($S=1.0$
day) to the Tohoku-Oki earthquake (March 11, 2011; dashed line). The almost
overlapping red curve indicates the theoretical ETAS cumulative function,
equation (\protect\ref{equ4}), and the extension to the rest of the period until April 2012.
The inset rectangle magnifies the cumulative curve for the extrapolated
period.}\label{fig3}
\end{figure}

An analysis of the 2008 earthquake aftershock sequence is shown in
Figure~\ref{fig3}. Here the ETAS model is fitted to the period from one day
after the main
shock until the M9.0 earthquake. The estimated intensity is then
extrapolated to span an additional year. The change point at the M9.0
earthquake is substantial, decreasing the total \textit{AIC} by $28.5$, showing a
relative quiescence afterward. The penalty quantity $q$ in the
$\mathit{AIC}_{12}$ of equation (\ref{equ9}) equals zero because the change
point is
given by the information outside of the aftershock data, hence, $\Delta
\mathit{AIC} = -28.5$. Therefore, the occurrence of the Tohoku-Oki
earthquake is a significant change point.

Hereafter, the data set becomes very difficult for conventional ETAS
analysis. The earthquake swarm near Lake Inawashiro began March 18,
2011, a
week after the M9.0 earthquake in region~B (Figure~\ref{fig1}). Seismic activity in
this area was very low before the M9.0 event. The swarm mostly
consisted of
small earthquakes with magnitudes less than $3.0$. The largest earthquake in
this cluster, an earthquake of~M4.6, occurred $50$ days after the M9.0
earthquake, and its aftershock sequence seemed to decay normally.

%
%
\begin{figure}

\includegraphics{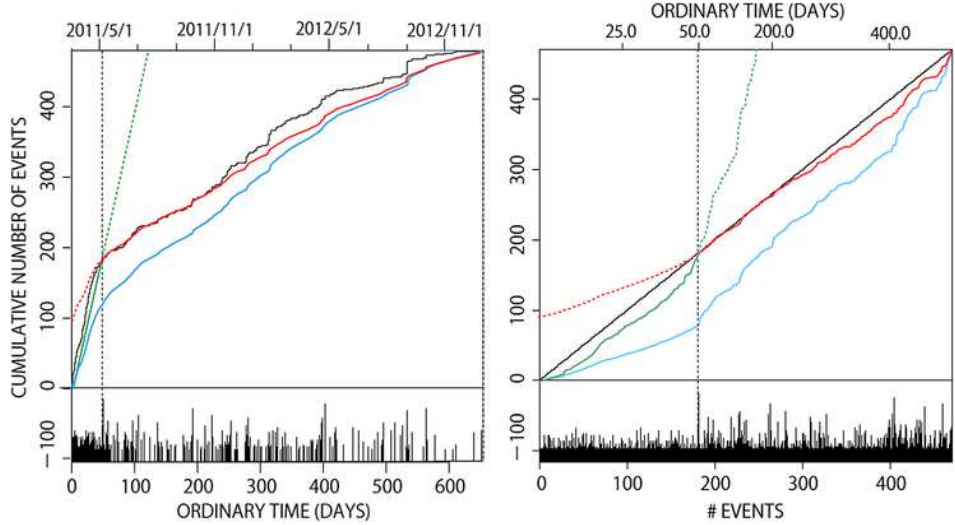}

\caption{Stationary and two-stage ETAS models fitted to region B.
The ETAS model is fitted to the entire period from March 18, 2011, to the
end of 2012 with the preliminary period of the first 0.1 days (blue line),
the period before the M4.6 event ($t = 49.8$ days) (green solid line) then
extrapolated forward (green dashed line), and the period after the M4.6
event (red solid line) then extrapolated backward (red dashed line). The
black curve shows the cumulative number of observed earthquakes. The left
panel plots these against ordinary time, whereas the right panel plots
these against the number of earthquakes.}\label{fig4}
\end{figure}
%
%
\begin{table}
\tabcolsep=0pt
\caption{The ETAS parameters of region B fitted to \textup{(a)} the entire
period, \textup{(b)} and \textup{(c)} before the change point, and \textup{(d)} after the change
point. Their standard errors are in parentheses. The improvement of the
two-stage ETAS model relative to the stationary ETAS model is $\Delta \mathit{AIC}
= (422.9 - 118.3) - 442.8 = -138.2$. The MLE for the change point is
$t = 49.8$, which coincides with the time just before the M4.6.
The threshold magnitude is $M_{z} = 2.5$. Numbers are
rounded to three significant digits}\label{tab2}
\begin{tabular*}{\tablewidth}{@{\extracolsep{\fill}}@{}ld{2.9}d{2.9}d{2.9}d{2.9}d{2.9}d{4.1}@{}}
\hline
\textbf{Period} & \multicolumn{1}{c}{$\bolds{\mu}$} & \multicolumn{1}{c}{$\bolds{K_{0}}$}
                & \multicolumn{1}{c}{$\bolds{c}$}   & \multicolumn{1}{c}{$\bolds{\alpha}$}
                & \multicolumn{1}{c}{$\bolds{p}$}   & \multicolumn{1}{c@{}}{$\bolds{\mathit{AIC}}$}\\
\hline
(a) The whole & 9.77 \times10^{-2} & 6.54 \times10^{-2} & 9.64 \times10^{-4} & 0.215 & 0.900 & 442.8 \\
period & (7.81 \times10^{-2}) & (2.37 \times10^{-2}) & (6.35\times10^{-4}) & (9.77 \times10^{-2}) & (9.84 \times10^{-3}) &
\\[3pt]
(b) Before  & 1.41 & 1.05 \times10^{-1} & 8.52 \times10^{-2} & 3.06 \times10^{-15} & 1.00 & -103.0\\
change point  & (3.39 \times10^{-1}) & (6.99 \times10^{-2}) & (1.09 \times10^{-1}) & (9.35 \times10^{-1}) & & \\
with fixed\\
$p = 1.0$
\\[3pt]
(c) Before  & 1.27 & 2.12 \times10^{+ 11} & 1.04 \times 10^{+1} & 2.25 \times10^{-12} & 1.13 \times10^{+1} & -118.3\\
change point  & (5.52 \times10^{-1}) & (4.71) & (3.81 \times 10^{-1}) & (1.03) & (2.31 \times10^{-1}) & \\
without\\
fixed $p$\\[3pt]
(d) After  & 6.58 \times10^{-2} & 3.58 \times10^{-2} & 7.11 \times10^{-5} & 0.912 & 0.945 & 422.9\\
change point & (1.43 \times10^{-1}) & (1.90 \times10^{-2}) & (1.01\times10^{-3}) & (1.10 \times10^{-1}) & (1.87 \times10^{-1}) &\\
\hline
\end{tabular*}
\end{table}

First, the stationary ETAS model is applied to the whole period. The
theoretical cumulative function (solid light blue curves, Figure~\ref{fig4}) is
biased below from the empirical cumulative function, indicating a
substantial misfit. Hence, the two-stage ETAS model is applied to the data
to search the MLE for a change point. Table~\ref{tab2} lists the estimated
parameters and \textit{AIC} values. The change-point analysis (cf. Section~\ref{sec22})
implies that the MLE of the change point is at $t = 49.8$ days from the
beginning of this cluster, which coincides with the time just before the
M4.6 earthquake occurred. The two-stage ETAS model with this change point
improves the \textit{AIC} by $138.2$ (see Table~\ref{tab2}). The first-stage ETAS model before
the change point, with a fixed parameter $p = 1.0$, still displays a large
deviation from the ideal fit (cf. the solid green curve in Figure~\ref{fig4}). The
magnitude sensitivity parameter $\alpha$ becomes very small relative to
that of the second-stage ETAS model. Such a small value implies that almost
all earthquakes in the first stage occurred independently to preceding
magnitudes (i.e., close to a Poisson process), and can be mostly attributed
to the average $\mu$ rate of the background seismicity. The first stage
$\mu$ rate is two orders of magnitude higher than the second stage rate.

If $p$ is not fixed, the estimated $K_{0}$, $c$ and $p$ have extremely
large values for a normal earthquake sequence while $\alpha$ approaches
zero. Consequently, the model is again approximate to a nonstationary
Poisson process, characterizing the sequence as a swarm, with an \textit{AIC}
smaller than that of the $p = 1.0$ scenario. The large discrepancies
between the estimated parameter values between (b) and (c) in Table~\ref{tab2}
suggest that the stationary ETAS model is not well defined for this
particular earthquake sequence in the first period before the change point.
The standard errors for the parameter $\alpha$ are multiple orders of
magnitude greater than those of the estimates themselves. The narrow
magnitude range makes it difficult for the model to distinguish the effects
of $K_{0}$ and $\alpha$, causing a trade-off between these two parameters,
thus providing inaccurate estimations. For the case without a fixed $p$,
the aftershock productivity $K_{0}$ becomes extremely small in compensation
for the small $\alpha$ estimate.

After the change-point time of the M4.6 earthquake, the ETAS model fits
considerably well for several months. Then, a deviation becomes noticeable
relative to the solid red cumulative curve in Figure~\ref{fig4}. From these
observations, it is concluded that the M4.6 earthquake has reduced swarm
activity and that decaying normal aftershock type activity has dominated.

\subsection{Comparison of the nonstationary models}\label{sec32} In
this section the
proposed nonstationary models and methods outlined in Sections~\ref{sec24}--\ref{sec26} are
applied to the same data from region B near Lake Inawashiro. To replicate
the transient nonstationary activities in this particular region, we use
the seismic activity in the larger polygonal region in Figure~\ref{fig1} for the
period before the M9.0 earthquake (MLEs are shown in Figure~\ref{fig2}). Such a
reference model represents a typical seismicity pattern over a wide region
throughout the period, and therefore represents a robust estimate against
the inclusion of local and transient anomalies.

By fixing the reference parameters $c, \alpha$ and $p$, both in the
stationary and two-stage ETAS models, $\mu$ and $K_{0}$ are estimated for
events from region B after the M9.0 event, with a magnitude $M\geq 2.5$. Table~\ref{tab3} summarizes the re-estimated parameters, together with
the corresponding \textit{AIC} values. The \textit{AIC} improvement of the two-stage ETAS
model is $126.2$.

%
%
\begin{table}
\tabcolsep=0pt
\caption{Reference parameters adjusted to the data from region B
and the parameters of the present two-stage ETAS model (standard errors in
parentheses) with fixed $c$, $\alpha$ and $p$ of the reference model (standard
errors in brackets), with their \textit{AIC} values. The improvement of the
two-stage ETAS model relative to the present stationary ETAS model is
$\Delta \mathit{AIC} = 434.7 - 95.4 - 465.5 = -126.2$. Also, the improvement of the
present two-stage ETAS model relative to the stationary ETAS model in
Table~\protect\ref{tab2} is as follows: $\Delta \mathit{AIC} = 434.7 - 95.4 - 442.8 = -103.5$.
The change point is at $t = 49.8$, corresponding to the time just before the
M4.6 earthquake. The threshold magnitude $M_{z} = 2.5$.
Numbers are rounded to three significant digits}\label{tab3}
\begin{tabular*}{\tablewidth}{@{\extracolsep{\fill}}@{}ld{2.9}d{2.9}d{3.9}d{2.9}d{2.9}d{3.1}@{}}
\hline
\textbf{Period} & \multicolumn{1}{c}{$\bolds{\mu}$} & \multicolumn{1}{c}{$\bolds{K_{0}}$}
                & \multicolumn{1}{c}{$\bolds{c}$}   & \multicolumn{1}{c}{$\bolds{\alpha}$}
                & \multicolumn{1}{c}{$\bolds{p}$}   & \multicolumn{1}{c@{}}{$\bolds{\mathit{AIC}}$}\\
\hline
(a) The whole & 1.92 \times10^{-1} & 2.49 \times10^{-2} & 6.02 \times10^{-3} & 2.03 & 1.11 & 465.5\\
period & (3.58 \times10^{-2}) & (5.82 \times10^{-3}) & {[}2.50 \times10^{-3}{]} & {[}1.27 \times10^{-2}{]} & {[}5.44 \times10^{-3}{]} &
\\[3pt]
(b) Before the & 3.31 & 6.77 \times10^{-3} & \multicolumn{1}{c}{$602 \times10^{-3}$\phantom{.00}} & 2.03 & 1.11 & -95.4\\
change point & (1.04 \times10^{-1}) & (3.27 \times10^{-3}) & & &&
\\[3pt]
(c) After the & 1.95 \times10^{-1} & 1.56 \times10^{-2} & 6.02 \times10^{-3} & 2.03 & 1.11 & 434.7\\
change point & (2.99 \times10^{-2}) & (6.41 \times10^{-3}) & & & \\
\hline
\end{tabular*}
\end{table}

Next, we have applied the nonstationary ETAS models listed in Table~\ref{tab1}, with
and without a change point taken into consideration, using the reference
parameters in the first row of Table~\ref{tab3}. Here, if a change point of M4.6 at
the time $t = 49.8$ days occurs, we propose a very small fixed value such
as that described in Section~\ref{sec25}.

%
%
\begin{figure}[t]

\includegraphics{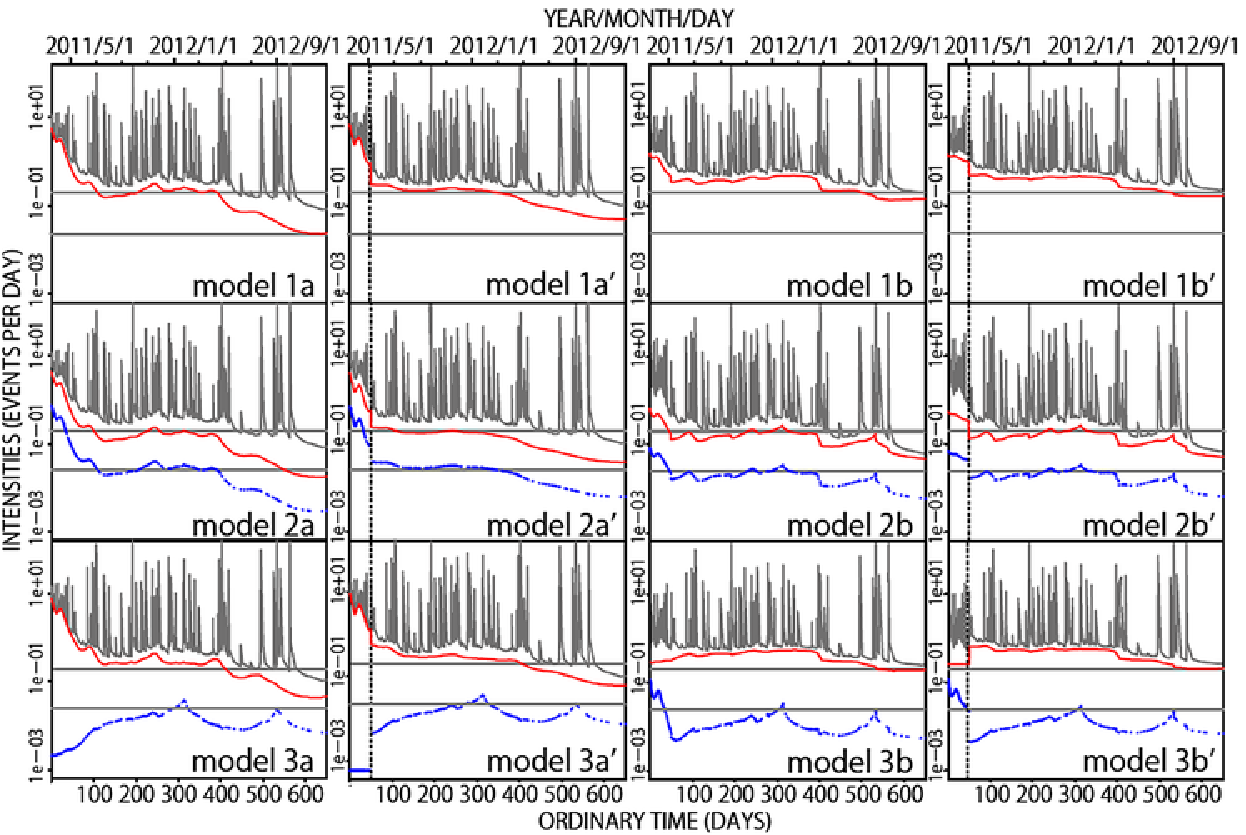}

\caption{Various inversion results of all considered models for the
data from region B. The model numbers correspond to those of Table~\protect\ref{tab1}, and
the models with prime ($'$) correspond to those that include a change
point. The background rates $\mu (t)$ are shown in red connected lines, and
the productivity $K_{0}(t)$ is shown in blue dots at earthquake
occurrence times. The gray spiky curves represent the conditional intensity
rates $\lambda (t | H_{t})$. The above three functions
are plotted on a logarithmic scale. The upper and lower gray horizontal
lines represent the reference parameters $\mu$ and $K_{0}$,
respectively (see Table~\protect\ref{tab2}). The vertical dashed line shows the change-point
time, $t = 4$ days elapsed from March~18, 2011. The horizontal
axis indicates days elapsed.}\label{fig5}
\end{figure}

Figure~\ref{fig5} shows all of the inversion results (maximum posterior estimates)
for a total of 12 models. The $\Delta\mathit{ABIC}$ values of the
corresponding models are given in Table~\ref{tab4}. Models with the change point
outperform corresponding models without the change point. This highlights
the significance of jumps at the change point. Such improvements via jumps
are smaller between corresponding models with constraints on the
transformed time. This is because those models already present jumps or
sharp changes to some extent in the target parameters even without setting
change points, due to the expanded transformed time during the dense event
period after the M4.6 event in ordinary time. Results also show that models
with constraints on ordinary time yield better results than those with the
transformed time. This is probably because the data set only contains
gradually changing parameters except at the change point.

\begin{figure}

\includegraphics{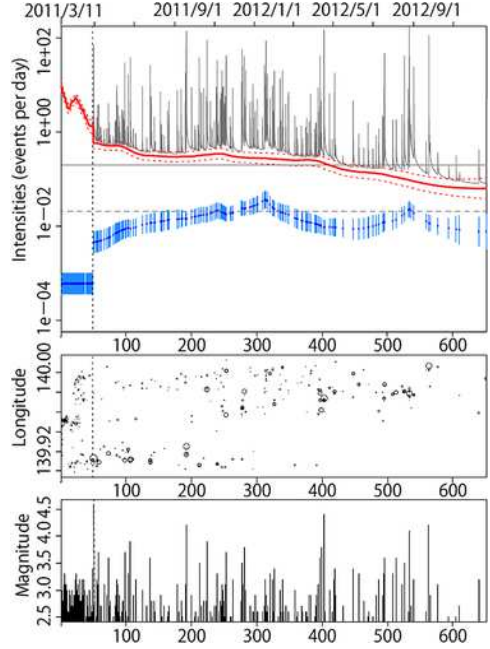}

\caption{The selected best-fitted model 3(a$'$) and errors of the
inversion solutions. The background rate $\mu$(t) is shown in solid red,
with $\mbox{one}-\sigma$ error bounds in red dashed lines. $K_{0}(t)$ is
shown in blue dots with $\mbox{one}-\sigma$ error bars at the occurrence times.
The gray spiky curve represents the variation of the intensity rates
$\lambda(t | H_{t})$. All of the above estimates are
plotted on a logarithmic scale. The solid gray horizontal line represents
the reference $\mu$ value, and the horizontal dashed line represents the
reference $K_{0}$ value (see Table~\protect\ref{tab2}). The horizontal axis is the
elapsed days from March 18, 2011. The vertical dashed line shows the change
point $t = 49.8$ elapsed days. The middle panel displays the
longitudes versus the elapsed times of the earthquake occurrences in region
B. The diameters of the circles are proportional to the earthquake
magnitudes. The bottom panel shows magnitudes of earthquakes versus the
ordinary elapsed times in days.}\label{fig6}
\end{figure}

The smallest $\Delta\mathit{ABIC}$ is achieved by model 3(a$'$) in
which both $q_{\mu} (t)$ and $q_{K}(t)$ are nonstationary on the smoothness
constraints under ordinary time, with a jump at the time of the M4.6
earthquake. Figure~\ref{fig6} shows variations of the background and productivity
rates in the selected nonstationary model. These variations suggest that
the intensity of aftershock productivity $K_{0} (t) ( = K_{0}
q_{K}(t))$ is
extremely low during early periods of earthquake swarms until the M4.6
earthquake occurs; meanwhile, the background seismicity $\mu(t) ( = \mu
q_{\mu} (t))$ changes at a high rate. Therefore, the total seismicity
$\lambda_{\theta} (t | H_{t})$ in that period is similar to a nonstationary
Poisson process with intensity rates $\mu(t)$ of the background activity.
After the M4.6 earthquake occurred, the $\mu(t)$ rate gradually decreased
while $K_{0} (t)$ increased. These changes are roughly approximated by the
estimated two-stage ETAS model in Table~\ref{tab3}, in which $\mu$ before the change
point is higher, while $K_{0}$ is lower than those after the change
point.

%
\begin{table}[b]
\tabcolsep=0pt
\caption{$\Delta\mathit{ABIC}$ value of each model defined in equation
(\protect\ref{equ25}). The underlined model has the smallest value. The prime ($'$)
indicates the models that further assume a change point at $t = 49.8$, the
time when the M4.6 earthquake occurred}\label{tab4}
\begin{tabular*}{\tablewidth}{@{\extracolsep{\fill}}@{}lcccc@{}}
\hline
\textbf{Models} & \textbf{a} & \textbf{a$\bolds{'}$} & \textbf{b} & \textbf{b}$\bolds{'}$\\
\hline
{1} & $-170.0$ & $-177.2$ & $-132.4$ & $-$134.1\\
{2} & $-175.3$ & $-180.1$ & $-136.1$ & $-$137.2\\
{3} & $-250.1$ & $-$\underline{260.8} & $-148.1$ & $-$151.5\\
\hline
\end{tabular*}
\end{table}
%

\begin{figure}

\includegraphics{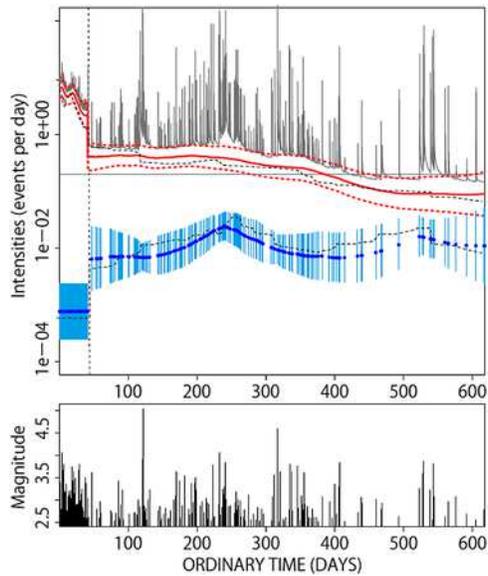}

\caption{The maximum a posteriori (MAP) solution of a synthesized
data set by the estimated model~3(\textup{a}$'$) (shown in Figure~\protect\ref{fig6}) with the same
reference parameters in Table~\protect\ref{tab2}. The re-estimated parameters $\mu(t)$ and
$K_{0}(t)$ are shown in red and blue curves, respectively, with
two-fold error bounds. The upper and lower dashed black curves represent
the true $\mu(t)$ and $K_{0}(t)$ (same as those in Figure~\protect\ref{fig6}),
respectively.}\label{fig7}
\end{figure}

If the $\Delta\mathit{ABIC}$ of model 3(a$'$) in Table~\ref{tab4} and
$\Delta
\mathit{AIC}$ of the two-stage ETAS models in Tables~\ref{tab2} and \ref{tab3} are compared
[\citeauthor{Aka85} (\citeyear{Aka85,Aka87})], the former model
displays a much better fit,
with a
difference of more than 130. This indicates that the specific details of
transient variations in model~3(a$'$) appear to be substantial. Model
3(a$'$) further shows that the background $\mu(t)$ rate decreased
after about $t = 400$ days, indicating that the swarm component of the
seismicity decreased. To demonstrate the reproducibility of the detailed
variations with the similar data sets, Figure~\ref{fig7} shows the re-estimated model
3(a$'$) utilizing the same optimization procedure from simulated data
in the estimated model 3(a$'$) in Figure~\ref{fig6}. See the \hyperref[sec6]{Appendix} for more
details.

%
%

\begin{figure}

\includegraphics{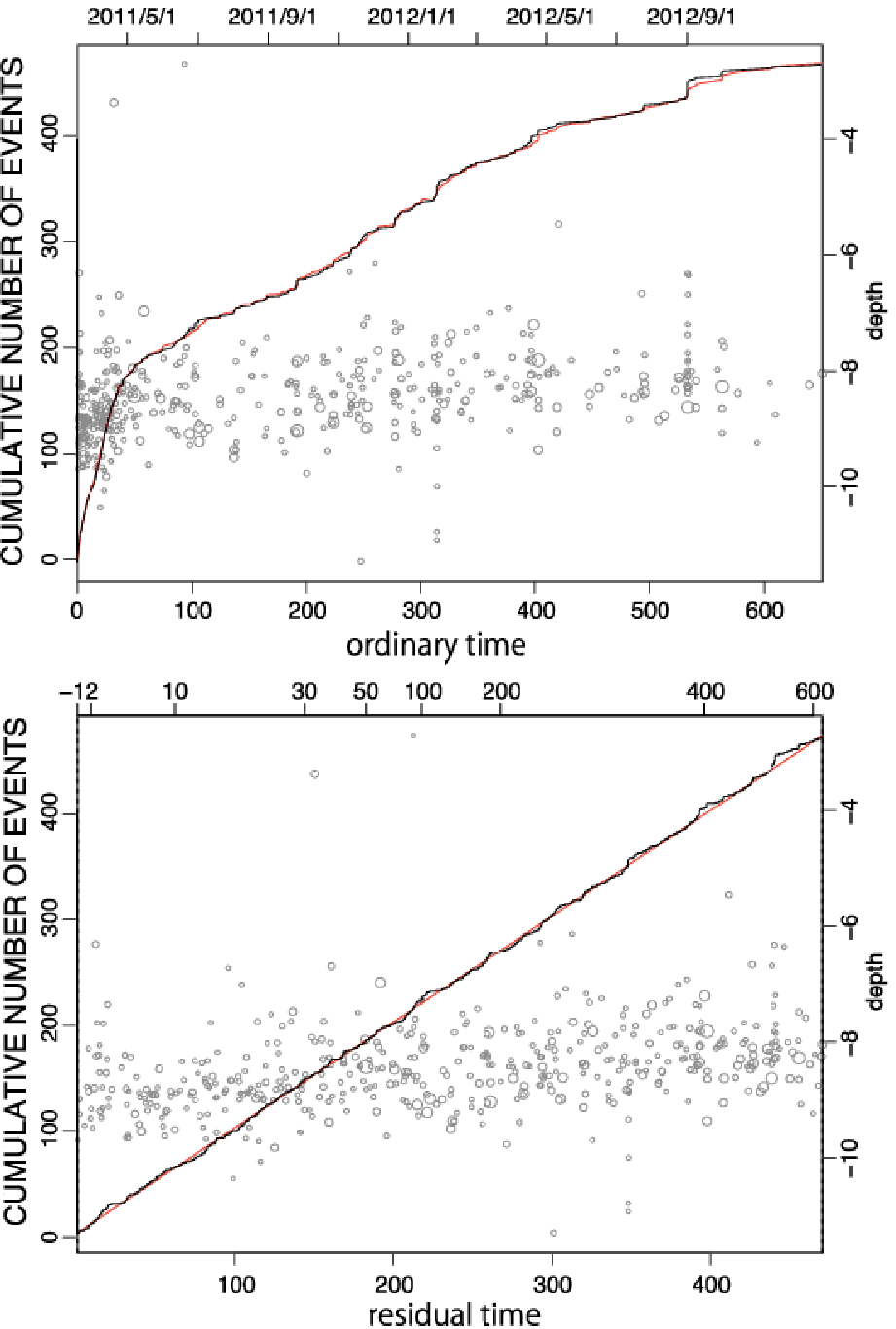}

\caption{Estimated cumulative number of events by model 3(\textup{a}$'$) (red
curves) and the observed number of events (black curve) for the ordinary
time (top panel) and residual time (bottom panel). Gray circles show the
depths of the swarm events versus the corresponding time.}\label{fig8}
\end{figure}

The model's performance is graphically examined by plotting the estimated
cumulative number of events (\ref{equ4}) to compare with the observed events in
Figure~\ref{fig8}, which shows that the observed events become almost a stationary
Poisson process, although a few clustering features remain.

%

\begin{figure}

\includegraphics{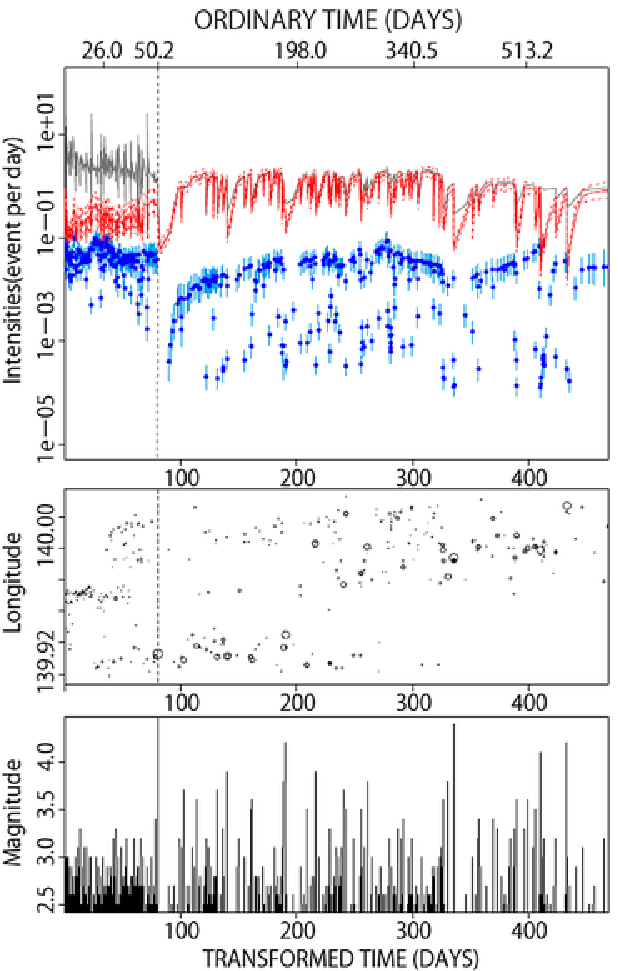}

\caption{Variations of conditional intensity rates $\lambda(\tau|
H_{\tau})$, background rate $\mu(\tau)$ and aftershock
productivity $K_{0}(\tau)$ of model 3(\textup{b}$'$) versus the transformed
time $\tau$ of the reference ETAS model (the first row of Table~\protect\ref{tab3}). The
other details are the same as those in Figure~\protect\ref{fig6}.}\label{fig9}
\end{figure}

It is worthwhile to discuss why model 3(b$'$) with constraints under
the transformed time has a poorer fit than model 3(a$'$) with
constraints under ordinary time. The MAP estimate of model 3(b$'$) is
shown in Figure~\ref{fig9}, where the transformed time $\tau$ in this case is
defined in equations (\ref{equ4}) and (\ref{equ5}) using the reference ETAS model, the
parameter value of which is listed in the first row of Table~\ref{tab3}. Parameter
variations in the period after the change point are similar to those of the
overall best model 3(a$'$). Variations during the period before the
change point are different with higher $K_{0} (t)$ and lower $\mu(t)$.
However, in this particular application, the performance of model
3(b$'$) on the whole is inferior in terms of $\Delta\mathit
{ABIC}$ by
a difference of greater than $100$. This may be because the above mentioned
reference ETAS-based transformed time of the former period worked poorly,
unlike during the latter period.

%
%

Although the goodness of fit of model 3(b$'$) over the whole period
(particularly during the former period) is not quite satisfactory, it is
worthwhile to examine the changes of $\mu( \tau)$ and $K_{0} (\tau)$
during the latter period in Figure~\ref{fig9}. The conditional intensity rate
$\lambda_{\theta} (t | H_{t})$, background rate $\mu( \tau)$ and
aftershock productivity rate $K_{0} (\tau)$ rapidly decrease not only after
the M4.6 earthquake but also after relatively large earthquakes. On such
sharp drops, there is a technical but simple explanation. Models in
Table~\ref{tab4}
with smoothness constraints on the transformed time are sensitive to
catalog incompleteness during small time intervals after large earthquakes.
In other words, a substantial number of small earthquakes that occur
immediately after a large earthquake are missing in the earthquake catalog
[e.g., \citet{OGAKAT06}; \citet{OMIetal13}]. Present results
suggest that the smoothing on the transformed time can be used as a
supplemental tool to check catalog completeness. The time transformation
stretches out ordinary time where the intensity rate is high and, hence,
transforming the smoothed parameters back to ordinary time can result in
sharp changes. This type of constraint can be useful for different
applications in which occasional rapid changes are expected.\looseness=-1

\subsection{Seismological complements and implications of the results}\label{sec33} Used
as a reference model, the polygonal region in Figure~\ref{fig1} is known to have a
similar seismicity pattern with similar focal mechanisms under the
west--east compressional tectonic field, as described in \citet{TERMAT}
and \citeauthor{TODLIAROS11} (\citeyear{TODLIAROS11,TODSTEJIA11}). For example, earthquakes have
mostly north--south strike angles and west--east directional thrust faults
in this region. This pattern can also be seen in the configurations of
active fault systems on the surface.

In the above sections, the estimation procedures of the models presented
here have been illustrated with a data set that includes a cluster of swarm
earthquakes triggered by the March 11, 2011 M9.0 Tohoku-Oki earthquake.
Swarm activity in this region seems to be triggered by surface waves
emitted from the M9.0 source, and has been studied by \citet{TERHASMAT13}
using the seismological theory and methods used in \citet{TERMILDEI12}. Here they attribute swarm activity to the weakening of the fault
via an increase of pore fluid pressure caused by the dynamic triggering
effect due to surface waves of the Tohoku-Oki rapture. Thus, the initially
very high and then decreasing rate of $\mu(t)$ reflects changes in fault
strength, probably due to the intrusion and decrease in pore fluid
pressure. The analyses presented here support the quantitative,
phenomenological evidence of fault weakening via the intrusion of water
into the fault system in earlier periods [\citeauthor{TERMILDEI12} (\citeyear{TERMILDEI12,TERHASMAT13})].
Similarly, by monitoring swarm activity, this nonstationary model can be
expected to make quantitative inferences of magma intrusions and draining
during volcanic activity.

The background seismicity parameter in the ETAS model is sensitive to
transient aseismic phenomena such as slow slips (quiet earthquakes) on and
around tectonic plate boundaries [\citet{LLEMCGOGA09}, \citet{OKUIDE11}]. This could possibly link a given swarm activity to the weakening
of interfaces. Changes in the pore fluid pressure, for example, alter the
friction rate of fault interfaces, thereby changing the fault strength.
Hence, monitoring the changes in background seismicity has the
potential to
detect such aseismic events.

Changes in the aftershock productivity $K_{0}$, on the other hand, appear
to depend on the locations of earthquake clusters and appear to vary among
clusters where secondary aftershocks are conspicuous. The aftershock
productivity $K_{0}$ therefore reflects the geology around faults rather
than the changes in stress rate. The application of the space--time ETAS
model with location-dependent parameters [e.g., \citet{OgaKatTan03},
\citeauthor{OGA04}
(\citeyear{OGA04,OGA11N2})] reveals that the $K_{0}$ function varies
(i.e., location
sensitive) unlike other parameters. Still, the task remains to confirm the
link between the changes in ETAS parameters and physical processes
happening on and around faults.

\section{Conclusions and discussion}\label{sec4} There are many
examples in seismology
in which different authors have obtained differing inversion results for
the same scientific phenomenon. These differences are attributed to the
adoption of different priors for the parameters of a given model. Scenarios
in this study have the same problem and are highlighted in Figure~\ref{fig5}. Model
parameters in this study are estimated by maximizing the penalized log
likelihood, which is intrinsically nonlinear. Besides adjusting the weights
in the penalty (namely, hyperparameters of a prior distribution), it is
necessary to compare the adequacy of different penalties (prior
distributions) associated with the same likelihood function. For these
purposes, we have proposed the objective procedure using $\Delta \mathit{ABIC}$ and $\Delta \mathit{AIC}$.

A suitable ETAS model [equation~(\ref{equ2})] is first established with MLE as the
reference predictive model to monitor future seismic activity and to detect
anomalous seismic activity. Sometimes, transient activity starts in a
region with very low seismicity. In such a case, it is both practical and
applicable to use a data set from a wider region to estimate the stable and
robust parameter values of $c$, $\alpha$ and $p$ in the ETAS model
[equation~(\ref{equ2})]. Then, the competing nonstationary ETAS models in equation
(\ref{equ10}) are fitted together with constraint functions in equations (\ref{equ11})~and~(\ref{equ12}) using either ordinary time or transformed time to penalize the
time-dependent parameters in the models. The corresponding Bayesian models
include a different prior distribution of the anomaly factor coefficients
$q_{\mu} (\cdot)$ and $q_{K}(\cdot)$, which are functions of either the ordinary
time $t$ [models 1(a)--3(a) in Table~\ref{tab1}] or the transformed
time $\tau$ in the reference ETAS model [models~\mbox{1(b)--3(b)}].
Furthermore, models in which the anomaly functions involve a discontinuity
[models 1(a$'$)--3(a$'$) and 1(b$'$)--3(b$'$)] are considered. Using the
$\Delta \mathit{ABIC}$ value, the goodness-of-fit performances of all of the
different models are summarized in Table~\ref{tab4}. Among the competing models,
model~3(a$'$) attained the smallest $\Delta \mathit{ABIC}$ value, and it is
therefore concluded that this model provides the best inversion result for
this particular data set.

Thus, changes in background seismicity $\mu$ and/or aftershock productivity
$K_{0}$ of the ETAS model can be monitored. The background seismicity rate
in the ETAS models represents a portion of the occurrence rate due to
external effects that are not included in the observed earthquake
occurrence history in the focal region of interest. Therefore, changes in
the background rate have been attracting the interest of many researchers
because such changes are sometimes precursors to large earthquakes. The
declustering algorithms [e.g., \citet{REA85}, \citeauthor{ZhuOgaVer02} (\citeyear{ZhuOgaVer02,ZHUOGAVER04})] have been adopted to determine the background seismicity by
stochastically removing the clustering components depending on the
ratio of
the background rate to the whole intensity at each occurrence time. The
change-point analysis and nonstationary models presented in this study,
however, objectively serve a more quantitatively explicit way to approach
this task.

The case where the other three parameters $c, \alpha$ and $p$ in equation
(\ref{equ2}) also vary with time was not examined in this study. For example, in
Figure~\ref{fig8}, we have seen that the best model in our framework does not
capture all of the clustering events but misses a few small clusters, which
suggests the time dependency of the parameters. For another example, we
have seen the effect of missing earthquakes in Figure~\ref{fig7}, suggesting that
parameter $c$ may depend on the magnitude of the earthquake, leading to a
significant correlation between $c$ and $p$. Furthermore, in Section~\ref{sec23},
it is mentioned that $K_{0}$ is correlated with the parameter $\alpha$.
Unstable estimations of $K_{0}$ and the $\alpha$ value in the swarm period
before the M4.6 earthquake can be seen in Table~\ref{tab2}, during which period most
of the magnitudes are between 2.5 and 3. This is another reason why the
$\alpha$ value is fixed by the corresponding reference parameter $\alpha$
when the nonstationary models are applied. Owing to the linearly
parameterized coefficients of the functions $q_{\mu}$ and $q_{K}$ in
equation (\ref{equ10}), the maximizing solutions of the penalized log-likelihood
function [equation~(\ref{equ16})], in spite of the high dimension, can be obtained
uniquely and stably by fixing the three parameters $c, \alpha$ and $p$.

\begin{appendix}
\section*{Appendix: Synthetic test of reproducibility of nonstationary patterns}\label{sec6}

We tested our method with synthetic data sets to check if both $\mu(t)$
and $K_{0} (t)$ can be reproduced by simulated data sets that are similar
to observed data sets. We used the reference parameter set (Table~\ref{tab2}) with
the best estimated $\mu(t)$ and $K_{0} (t)$ of model 3(a$'$).

The magnitude sequence of the synthetic data was generated on the basis of
the Gutenberg--Richter law with a $b$-value of the original data set
($b =
1.273$). In other words, the magnitude of each earthquake will
independently obey an exponential distribution such that $f (M) = \beta\exp\{- \beta(M - M_{c})$, $M \geq M_{c}$, where $\beta= b\ln10$, and
$M_{c} = 2.5$ is the magnitude value above which all earthquakes are
detected.

The thinning method [\citeauthor{OGA81} (\citeyear{OGA81,OGA98})] is
adopted for data
simulation. A~total of 470 events were simulated with a threshold magnitude of 2.5. Model
3(a$'$) was then fitted to the simulated data sets, with a change point
at the same time as the original data (between the 182nd and 183rd event).
Results are shown in Figure~\ref{fig7}; the estimated $\mu(t)$ and $K_{0} (t)$
appear to be similar to the original $\mu(t)$ and $K_{0} (t)$ in
Figure~\ref{fig6},
respectively, within a $2 \sigma$ error.
\end{appendix}



\section*{Acknowledgments}\label{sec5}
We are grateful to the Japan Meteorological Agency (JMA), the National
Research Institute for Earth Science and Disaster Prevention (NIED) and
the universities for the hypocenter data. We used the TSEIS visualization
program package [\citet{TSU96}] for the study of hypocenter data.


%
%

\printaddresses

\begin{thebibliography}{76}
\bibitem[\protect\citeauthoryear{Adelfio and Ogata}{2010}]{AdeOga10}
%
\begin{barticle}[mr]
\bauthor{\bsnm{Adelfio},~\bfnm{Giada}\binits{G.}} \AND
\bauthor{\bsnm{Ogata},~\bfnm{Yosihiko}\binits{Y.}}
(\byear{2010}).
\btitle{Hybrid kernel estimates of space--time earthquake occurrence
rates using the epidemic-type aftershock sequence model}.
\bjournal{Ann. Inst. Statist. Math.}
\bvolume{62}
\bpages{127--143}.
\bid{doi={10.1007/s10463-009-0268-7}, issn={0020-3157}, mr={2577443}}
\end{barticle}
%
\bptok{imsref}%
\endbibitem



\bibitem[\protect\citeauthoryear{Akaike}{1973}]{Ak73}
\begin{bincollection}[mr]
\bauthor{\bsnm{Akaike},~\bfnm{H.}\binits{H.}}
(\byear{1973}).
\btitle{Information theory and an extension of the maximum likelihood principle}.
In \bbooktitle{Second {I}nternational {S}ymposium on {I}nformation {T}heory ({T}sahkadsor, 1971)}
\bpages{267--281}.
\bpublisher{Akad\'emiai Kiad\'o},
\blocation{Budapest}.
\bid{mr={0483125}}
\end{bincollection}
%
\bptok{imsref}%
\endbibitem

\bibitem[\protect\citeauthoryear{Akaike}{1974}]{Ak74}
\begin{barticle}[mr]
\bauthor{\bsnm{Akaike},~\bfnm{Hirotugu}\binits{H.}}
(\byear{1974}).
\btitle{A new look at the statistical model identification}.
\bjournal{IEEE Trans. Automat. Control}
\bvolume{AC-19}
\bpages{716--723}.
\bnote{System identification and time-series analysis}.
\bid{issn={0018-9286}, mr={0423716}}
\end{barticle}
%
\bptok{imsref}%
\endbibitem

\bibitem[\protect\citeauthoryear{Akaike}{1977}]{Ak77}
\begin{bincollection}[mr]
\bauthor{\bsnm{Akaike},~\bfnm{Hirotugu}\binits{H.}}
(\byear{1977}).
\btitle{On entropy maximization principle}.
In \bbooktitle{Applications of Statistics}
(\beditor{\bfnm{P.~R.}\binits{P.~R.}~\bsnm{Krishnaian}}, ed.)
\bpages{27--41}.
\bpublisher{North-Holland},
\blocation{Amsterdam}.
\bid{mr={0501456}}
\end{bincollection}
%
\bptok{imsref}%
\endbibitem

\bibitem[\protect\citeauthoryear{Akaike}{1980}]{Aka80}
%
\begin{bincollection}[mr]
\bauthor{\bsnm{Akaike},~\bfnm{Hirotugu}\binits{H.}}
(\byear{1980}).
\btitle{Likelihood and the {B}ayes procedure}.
In \bbooktitle{Bayesian Statistics ({V}alencia, 1979)}
(\beditor{\bfnm{J.~M.}\binits{J.~M.}~\bsnm{Bernardo}},
\beditor{\bfnm{M.~H.}\binits{M.~H.}~\bsnm{De Groot}},
\beditor{\bfnm{D.~V.}\binits{D.~V.}~\bsnm{Lindley}}
\AND
\beditor{\bfnm{A.~F.~M.}\binits{A.~F.~M.}~\bsnm{Smith}}, eds.)
\bpages{143--166}.
\bpublisher{Univ. Press, Valencia},
\blocation{Spain}.
\bid{mr={0638876}}
\bptnote{check related}%
\end{bincollection}
%
\bptok{imsref}%
\endbibitem

\bibitem[\protect\citeauthoryear{Akaike}{1985}]{Aka85}
%
\begin{bincollection}[mr]
\bauthor{\bsnm{Akaike},~\bfnm{Hirotugu}\binits{H.}}
(\byear{1985}).
\btitle{Prediction and entropy}.
In \bbooktitle{A Celebration of Statistics}
(\beditor{\bfnm{A.~C.}\binits{A.~C.}~\bsnm{Atkinson}}
\AND
\beditor{\bfnm{E.}\binits{E.}~\bsnm{Fienberg}}, eds.)
\bpages{1--24}.
\bpublisher{Springer},
\blocation{New York}.
\bid{mr={0816143}}
\end{bincollection}
%
\bptok{imsref}%
\endbibitem

\bibitem[\protect\citeauthoryear{Akaike}{1987}]{Aka87}
%
\begin{barticle}[mr]
\bauthor{\bsnm{Akaike},~\bfnm{Hirotugu}\binits{H.}}
(\byear{1987}).
\btitle{Factor analysis and {AIC}}.
\bjournal{Psychometrika}
\bvolume{52}
\bpages{317--332}.
\bid{doi={10.1007/BF02294359}, issn={0033-3123}, mr={0914459}}
\end{barticle}
%
\bptok{imsref}%
\endbibitem

\bibitem[\protect\citeauthoryear{Balderama et~al.}{2012}]{Baletal12}
%
\begin{barticle}[mr]
\bauthor{\bsnm{Balderama},~\bfnm{Earvin}\binits{E.}},
\bauthor{\bsnm{Paik Schoenberg},~\bfnm{Frederic}\binits{F.}},
\bauthor{\bsnm{Murray},~\bfnm{Erin}\binits{E.}} \AND
\bauthor{\bsnm{Rundel},~\bfnm{Philip~W.}\binits{P.~W.}}
(\byear{2012}).
\btitle{Application of branching models in the study of invasive species}.
\bjournal{J. Amer. Statist. Assoc.}
\bvolume{107}
\bpages{467--476}.
\bid{doi={10.1080/01621459.2011.641402}, issn={0162-1459}, mr={2980058}}
\end{barticle}
%
\bptok{imsref}%
\endbibitem

\bibitem[\protect\citeauthoryear{Bansal and Ogata}{2013}]{BANOGA13}
%
\begin{barticle}[auto:STB|2014/06/18|12:29:53]
\bauthor{\bsnm{Bansal},~\bfnm{A.~R.}\binits{A.~R.}} \AND
\bauthor{\bsnm{Ogata},~\bfnm{Y.}\binits{Y.}}
(\byear{2013}).
\btitle{A non-stationary epidemic type aftershock sequence model for
seismicity prior to the December 26, 2004 M9.1 Sumatra-Andaman Islands
mega-earthquake}.
\bjournal{J. Geophys. Res.}
\bvolume{118}
\bpages{616--629}.
\bid{doi={10.1002/jgrb.50068}}
\end{barticle}
%
\bptok{imsref}%
\endbibitem

\bibitem[\protect\citeauthoryear{Chavez-Demoulina and Mcgillb}{2012}]{CHAMCG12}
%
\begin{barticle}[auto:STB|2014/06/18|12:29:53]
\bauthor{\bsnm{Chavez-Demoulina},~\bfnm{V.}\binits{V.}} \AND
\bauthor{\bsnm{Mcgillb},~\bfnm{J.~A.}\binits{J.~A.}}
(\byear{2012}).
\btitle{High-frequency financial data modeling using Hawkes processes}.
\bjournal{J. Bank. Financ.}
\bvolume{36}
\bpages{3415--3426}.
\end{barticle}
%
\bptok{imsref}%
\endbibitem

\bibitem[\protect\citeauthoryear{Daley and Vere-Jones}{2003}]{DALVER03}
%
\begin{bbook}[auto:STB|2014/06/18|12:29:53]
\bauthor{\bsnm{Daley},~\bfnm{D.}\binits{D.}} \AND
\bauthor{\bsnm{Vere-Jones},~\bfnm{D.}\binits{D.}}
(\byear{2003}).
\btitle{An Introduction to the Theory of Point Processes},
\bedition{2nd} ed.
\bpublisher{Springer},
\blocation{New York}.
\end{bbook}
%
\bptok{imsref}%
\endbibitem

\bibitem[\protect\citeauthoryear{Good}{1965}]{Goo65}
%
\begin{bbook}[mr]
\bauthor{\bsnm{Good},~\bfnm{Irving~John}\binits{I.~J.}}
(\byear{1965}).
\btitle{The Estimation of Probabilities. {A}n Essay on Modern
{B}ayesian Methods}.
\bpublisher{MIT Press},
\blocation{Cambridge, MA}.
\bid{mr={0185724}}
\end{bbook}
%
\bptok{imsref}%
\endbibitem

\bibitem[\protect\citeauthoryear{Good and Gaskins}{1971}]{GooGas71}
%
\begin{barticle}[mr]
\bauthor{\bsnm{Good},~\bfnm{I.~J.}\binits{I.~J.}} \AND
\bauthor{\bsnm{Gaskins},~\bfnm{R.~A.}\binits{R.~A.}}
(\byear{1971}).
\btitle{Nonparametric roughness penalties for probability densities}.
\bjournal{Biometrika}
\bvolume{58}
\bpages{255--277}.
\bid{issn={0006-3444}, mr={0319314}}
\end{barticle}
%
\bptok{imsref}%
\endbibitem

\bibitem[\protect\citeauthoryear{Hainzl and Ogata}{2005}]{HAIOGA05}
%
\begin{barticle}[auto:STB|2014/06/18|12:29:53]
\bauthor{\bsnm{Hainzl},~\bfnm{S.}\binits{S.}} \AND
\bauthor{\bsnm{Ogata},~\bfnm{Y.}\binits{Y.}}
(\byear{2005}).
\btitle{Detecting fluid signals in seismicity data through statistical
earthquake modeling}.
\bjournal{J. Geophys. Res.}
\bvolume{110}
\bpages{B5, B05S07}.
\end{barticle}
%
\bptok{imsref}%
\endbibitem

\bibitem[\protect\citeauthoryear{Hassan Zadeh and Sharda}{2012}]{HAS}
%
\begin{bmisc}[auto:STB|2014/06/18|12:29:53]
\bauthor{\bsnm{Hassan Zadeh},~\bfnm{A.}\binits{A.}}
\AND
\bauthor{\bsnm{Sharda},~\bfnm{R.}\binits{R.}}
(\byear{2012}).
\bhowpublished{Modeling brand post popularity in online social networks.
{Social Science Research Network}. Available at SSRN 2182711.}
\end{bmisc}
%
\bptok{imsref}%
\endbibitem

\bibitem[\protect\citeauthoryear{Hawkes}{1971}]{Haw71}
%
\begin{barticle}[mr]
\bauthor{\bsnm{Hawkes},~\bfnm{Alan~G.}\binits{A.~G.}}
(\byear{1971}).
\btitle{Spectra of some self-exciting and mutually exciting point processes.}
\bjournal{Biometrika}
\bvolume{58}
\bpages{83--90}.
\bid{issn={0006-3444}, mr={0278410}}
\end{barticle}
%
\bptok{imsref}%
\endbibitem

\bibitem[\protect\citeauthoryear{Hawkes and Adamopoulos}{1973}]{HAWADA73}
%
\begin{barticle}[auto:STB|2014/06/18|12:29:53]
\bauthor{\bsnm{Hawkes},~\bfnm{A.~G.}\binits{A.~G.}} \AND
\bauthor{\bsnm{Adamopoulos},~\bfnm{L.}\binits{L.}}
(\byear{1973}).
\btitle{Cluster models for earthquakes---regional comparisons}.
\bjournal{Bull. Int. Stat. Inst.}
\bvolume{45}
\bpages{454--461}.
\end{barticle}
%
\bptok{imsref}%
\endbibitem

\bibitem[\protect\citeauthoryear{Hawkes and Oakes}{1974}]{HawOak74}
%
\begin{barticle}[mr]
\bauthor{\bsnm{Hawkes},~\bfnm{Alan~G.}\binits{A.~G.}} \AND
\bauthor{\bsnm{Oakes},~\bfnm{David}\binits{D.}}
(\byear{1974}).
\btitle{A cluster process representation of a self-exciting process}.
\bjournal{J. Appl. Probab.}
\bvolume{11}
\bpages{493--503}.
\bid{issn={0021-9002}, mr={0378093}}
\end{barticle}
%
\bptok{imsref}%
\endbibitem

\bibitem[\protect\citeauthoryear{Herrera and Schipp}{2009}]{HERSCH09}
%
\begin{bincollection}[auto:STB|2014/06/18|12:29:53]
\bauthor{\bsnm{Herrera},~\bfnm{R.}\binits{R.}} \AND
\bauthor{\bsnm{Schipp},~\bfnm{B.}\binits{B.}}
(\byear{2009}).
\btitle{Self-exciting extreme value models for stock market crashes}.
In \bbooktitle{Statistical Inference, Econometric Analysis and Matrix Algebra}
\bpages{209--231}.
\bpublisher{Physica-Verlag HD}, \blocation{Heidelberg}.
\end{bincollection}
%
\bptok{imsref}%
\endbibitem

\bibitem[\protect\citeauthoryear{Japan Meteorological Agency}{2009}]{Age}
%
\begin{barticle}[auto:STB|2014/06/18|12:29:53]
\bauthor{\bsnm{Japan Meteorological Agency}}
(\byear{2009}).
\btitle{The Iwate-Miyagi Nairiku earthquake in 2008}.
\bjournal{Rep. Coord. Comm. Earthq. Predict}
\bvolume{81}
\bpages{101--131}.
\bnote{Available at \url
{http://cais.gsi.go.jp/YOCHIREN/report/kaihou81/03\_04.pdf}.}
\end{barticle}
%
\bptok{imsref}%
\endbibitem

\bibitem[\protect\citeauthoryear{Jordan, Chen and Gasparini}{2012}]{JORCHEGAS12}
%
\begin{barticle}[auto:STB|2014/06/18|12:29:53]
\bauthor{\bsnm{Jordan},~\bfnm{T.~H.}\binits{T.~H.}},
\bauthor{\bsnm{Chen},~\bfnm{Y.-T.}\binits{Y.-T.}} \AND
\bauthor{\bsnm{Gasparini},~\bfnm{P.}\binits{P.}}
(\byear{2012}).
\btitle{Operational earthquake forecasting. State of knowledge and
guidelines for utilization}.
\bjournal{Ann. Geophys.}
\bvolume{54}
\bpages{315--391}.
\end{barticle}
%
\bptok{imsref}%
\endbibitem

\bibitem[\protect\citeauthoryear{Kagan and Knopoff}{1987}]{KagKno87}
%
\begin{barticle}[pbm]
\bauthor{\bsnm{Kagan},~\bfnm{Y.~Y.}\binits{Y.~Y.}} \AND
\bauthor{\bsnm{Knopoff},~\bfnm{L.}\binits{L.}}
(\byear{1987}).
\btitle{Statistical short-term earthquake prediction}.
\bjournal{Science}
\bvolume{236}
\bpages{1563--1567}.
\bid{doi={10.1126/science.236.4808.1563}, issn={0036-8075},
pii={236/4808/1563}, pmid={17835741}}
\end{barticle}
%
\bptok{imsref}%
\endbibitem

\bibitem[\protect\citeauthoryear{Kendall}{1949}]{Ken49}
%
\begin{barticle}[mr]
\bauthor{\bsnm{Kendall},~\bfnm{David~G.}\binits{D.~G.}}
(\byear{1949}).
\btitle{Stochastic processes and population growth}.
\bjournal{J. Roy. Statist. Soc. Ser. B.}
\bvolume{11}
\bpages{230--264}.
\bid{issn={0035-9246}, mr={0034977}}
\end{barticle}
%
\bptok{imsref}%
\endbibitem

\bibitem[\protect\citeauthoryear{Kumazawa, Ogata and Toda}{2010}]{KUMOGATOD10}
%
\begin{barticle}[auto:STB|2014/06/18|12:29:53]
\bauthor{\bsnm{Kumazawa},~\bfnm{T.}\binits{T.}},
\bauthor{\bsnm{Ogata},~\bfnm{Y.}\binits{Y.}} \AND
\bauthor{\bsnm{Toda},~\bfnm{S.}\binits{S.}}
(\byear{2010}).
\btitle{Precursory seismic anomalies and transient crustal deformation
prior to the 2008 $M_w= 6.9$ Iwate-Miyagi Nairiku, Japan, earthquake}.
\bjournal{J.~Geophys. Res.}
\bvolume{115}
\bpages{B10312}.
\bid{doi={10.1029/2010JB007567}}
\end{barticle}
%
\bptok{imsref}%
\endbibitem

\bibitem[\protect\citeauthoryear{Laplace}{1774}]{Sti86}
%
\begin{barticle}[auto]
\bauthor{\bsnm{Laplace},~\bfnm{P.~S.}\binits{P.~S.}}
(\byear{1774}).
\btitle{Memoir on the probability of causes of events.
M\'emoires de math\'ematique et de physique, tome sixi\`eme
(English~translation by S.~M. Stigler, 1986)}.
\bjournal{Statist. Sci.}
\bvolume{1}
\bpages{364--378}.
\end{barticle}
%
\bptok{imsref}%
\endbibitem

\bibitem[\protect\citeauthoryear{Llenos, Mcguire and Ogata}{2009}]{LLEMCGOGA09}
%
\begin{barticle}[auto:STB|2014/06/18|12:29:53]
\bauthor{\bsnm{Llenos},~\bfnm{A.~L.}\binits{A.~L.}},
\bauthor{\bsnm{Mcguire},~\bfnm{J.~J.}\binits{J.~J.}} \AND
\bauthor{\bsnm{Ogata},~\bfnm{Y.}\binits{Y.}}
(\byear{2009}).
\btitle{Modeling seismic swarms triggered by aseismic transients}.
\bjournal{Earth Planet. Sci. Lett.}
\bvolume{281}
\bpages{59--69}.
\bid{doi={10.1016/j.epsl.2009.02.011}}
\end{barticle}
%
\bptok{imsref}%
\endbibitem

\bibitem[\protect\citeauthoryear{Lombardi, Cocco and Marzocchi}{2010}]{LOMCOCMAR10}
%
\begin{barticle}[auto:STB|2014/06/18|12:29:53]
\bauthor{\bsnm{Lombardi},~\bfnm{A.~M.}\binits{A.~M.}},
\bauthor{\bsnm{Cocco},~\bfnm{M.}\binits{M.}} \AND
\bauthor{\bsnm{Marzocchi},~\bfnm{W.}\binits{W.}}
(\byear{2010}).
\btitle{On the increase of background seismicity rate during the
1997--1998 Umbria--Marche, central Italy, sequence: Apparent variation
or fluid-driven triggering?}
\bjournal{Bull. Seismol. Soc. Amer.}
\bvolume{100}
\bpages{1138--1152}.
\end{barticle}
%
\bptok{imsref}%
\endbibitem

\bibitem[\protect\citeauthoryear{Lomnitz}{1974}]{LOM74}
%
\begin{bbook}[auto:STB|2014/06/18|12:29:53]
\bauthor{\bsnm{Lomnitz},~\bfnm{C.}\binits{C.}}
(\byear{1974}).
\btitle{Global Tectonic and Earthquake Risk}.
\bpublisher{Elsevier},
\blocation{Amsterdam}.
\end{bbook}
%
\bptok{imsref}%
\endbibitem

\bibitem[\protect\citeauthoryear{Mohler et~al.}{2011}]{Mohetal11}
%
\begin{barticle}[mr]
\bauthor{\bsnm{Mohler},~\bfnm{G.~O.}\binits{G.~O.}},
\bauthor{\bsnm{Short},~\bfnm{M.~B.}\binits{M.~B.}},
\bauthor{\bsnm{Brantingham},~\bfnm{P.~J.}\binits{P.~J.}},
\bauthor{\bsnm{Schoenberg},~\bfnm{F.~P.}\binits{F.~P.}} \AND
\bauthor{\bsnm{Tita},~\bfnm{G.~E.}\binits{G.~E.}}
(\byear{2011}).
\btitle{Self-exciting point process modeling of crime}.
\bjournal{J. Amer. Statist. Assoc.}
\bvolume{106}
\bpages{100--108}.
\bid{doi={10.1198/jasa.2011.ap09546}, issn={0162-1459}, mr={2816705}}
\end{barticle}
%
\bptok{imsref}%
\endbibitem

\bibitem[\protect\citeauthoryear{Ogata}{1978}]{Oga78}
%
\begin{barticle}[mr]
\bauthor{\bsnm{Ogata},~\bfnm{Yosihiko}\binits{Y.}}
(\byear{1978}).
\btitle{The asymptotic behaviour of maximum likelihood estimators for
stationary point processes}.
\bjournal{Ann. Inst. Statist. Math.}
\bvolume{30}
\bpages{243--261}.
\bid{doi={10.1007/BF02480216}, issn={0020-3157}, mr={0514494}}
\end{barticle}
%
\bptok{imsref}%
\endbibitem

\bibitem[\protect\citeauthoryear{Ogata}{1981}]{OGA81}
%
\begin{barticle}[auto:STB|2014/06/18|12:29:53]
\bauthor{\bsnm{Ogata},~\bfnm{Y.}\binits{Y.}}
(\byear{1981}).
\btitle{On Lewis' simulation method for point processes}.
\bjournal{IEEE Trans. Inform. Theory}
\bvolume{27}
\bpages{23--31}.
\end{barticle}
%
\bptok{imsref}%
\endbibitem

\bibitem[\protect\citeauthoryear{Ogata}{1985}]{OGA85}
%
\begin{bmisc}[auto:STB|2014/06/18|12:29:53]
\bauthor{\bsnm{Ogata},~\bfnm{Y.}\binits{Y.}}
(\byear{1985}).
\bhowpublished{Statistical models for earthquake occurrences and
residual analysis for point processes.
Research Memorandum No.~388 (21 May), The Institute of Statistical Mathematics,
Tokyo. Available at \url{http://www.ism.ac.jp/editsec/resmemo-e.html}.}
\end{bmisc}
%
\bptok{imsref}%
\endbibitem

\bibitem[\protect\citeauthoryear{Ogata}{1986}]{OGA86}
%
\begin{barticle}[auto:STB|2014/06/18|12:29:53]
\bauthor{\bsnm{Ogata},~\bfnm{Y.}\binits{Y.}}
(\byear{1986}).
\btitle{Statistical models for earthquake occurrences and residual
analysis for point processes}.
\bjournal{Mathematical Seismology}
\bvolume{1}
\bpages{228--281}.
\end{barticle}
%
\bptok{imsref}%
\endbibitem

\bibitem[\protect\citeauthoryear{Ogata}{1988}]{OGA88}
%
\begin{barticle}[auto:STB|2014/06/18|12:29:53]
\bauthor{\bsnm{Ogata},~\bfnm{Y.}\binits{Y.}}
(\byear{1988}).
\btitle{Statistical models for earthquake occurrences and residual
analysis for point processes}.
\bjournal{J. Amer. Statist. Assoc.}
\bvolume{83}
\bpages{9--27}.
\end{barticle}
%
\bptok{imsref}%
\endbibitem

\bibitem[\protect\citeauthoryear{Ogata}{1989}]{OGA89}
%
\begin{barticle}[auto:STB|2014/06/18|12:29:53]
\bauthor{\bsnm{Ogata},~\bfnm{Y.}\binits{Y.}}
(\byear{1989}).
\btitle{Statistical model for standard seismicity and detection of
anomalies by residual analysis}.
\bjournal{Tectonophysics}
\bvolume{169}
\bpages{159--174}.
\end{barticle}
%
\bptok{imsref}%
\endbibitem

\bibitem[\protect\citeauthoryear{Ogata}{1992}]{OGA92}
%
\begin{barticle}[auto:STB|2014/06/18|12:29:53]
\bauthor{\bsnm{Ogata},~\bfnm{Y.}\binits{Y.}}
(\byear{1992}).
\btitle{Detection of precursory relative quiescence before great
earthquakes through a statistical model}.
\bjournal{J. Geophys. Res.}
\bvolume{97}
\bpages{19845--19871}.
\end{barticle}
%
\bptok{imsref}%
\endbibitem

\bibitem[\protect\citeauthoryear{Ogata}{1998}]{OGA98}
%
\begin{barticle}[auto:STB|2014/06/18|12:29:53]
\bauthor{\bsnm{Ogata},~\bfnm{Y.}\binits{Y.}}
(\byear{1998}).
\btitle{Space--time point-process models for earthquake occurrences}.
\bjournal{Ann. Inst. Statist. Math.}
\bvolume{50}
\bpages{379--402}.
\end{barticle}
%
\bptok{imsref}%
\endbibitem

\bibitem[\protect\citeauthoryear{Ogata}{1999}]{OGA99}
%
\begin{barticle}[auto:STB|2014/06/18|12:29:53]
\bauthor{\bsnm{Ogata},~\bfnm{Y.}\binits{Y.}}
(\byear{1999}).
\btitle{Seismicity analysis through point-process modeling: A review}.
\bjournal{Pure Appl. Geophys.}
\bvolume{155}
\bpages{471--507}.
\end{barticle}
%
\bptok{imsref}%
\endbibitem

\bibitem[\protect\citeauthoryear{Ogata}{2001}]{Oga01}
%
\begin{barticle}[mr]
\bauthor{\bsnm{Ogata},~\bfnm{Yosihiko}\binits{Y.}}
(\byear{2001}).
\btitle{Exploratory analysis of earthquake clusters by likelihood-based
trigger models}.
\bjournal{J. Appl. Probab.}
\bvolume{38A}
\bpages{202--212}.
\bid{doi={10.1239/jap/1085496602}, issn={0021-9002}, mr={1915545}}
\end{barticle}
%
\bptok{imsref}%
\endbibitem

\bibitem[\protect\citeauthoryear{Ogata}{2004}]{OGA04}
%
\begin{barticle}[auto:STB|2014/06/18|12:29:53]
\bauthor{\bsnm{Ogata},~\bfnm{Y.}\binits{Y.}}
(\byear{2004}).
\btitle{Space--time model for regional seismicity and detection of
crustal stress changes}.
\bjournal{J. Geophys. Res.}
\bvolume{109}
\bpages{B03308}.
\bid{doi={10.1029/2003JB002621}}
\end{barticle}
%
\bptok{imsref}%
\endbibitem

\bibitem[\protect\citeauthoryear{Ogata}{2005}]{OGA}
%
\begin{barticle}[auto:STB|2014/06/18|12:29:53]
\bauthor{\bsnm{Ogata},~\bfnm{Y.}\binits{Y.}}
(\byear{2005}).
\btitle{Detection of anomalous seismicity as a stress change sensor}.
\bjournal{J. Geophys. Res.}
\bvolume{110}
\bpages{B05S06}.
\end{barticle}
%
\bptok{imsref}%
\endbibitem

\bibitem[\protect\citeauthoryear{Ogata}{2006a}]{OGA06N1}
%
\begin{barticle}[auto:STB|2014/06/18|12:29:53]
\bauthor{\bsnm{Ogata},~\bfnm{Y.}\binits{Y.}}
(\byear{2006}a).
\btitle{Seismicity anomaly scenario prior to the major recurrent
earthquakes off the East coast of Miyagi prefecture, northern Japan}.
\bjournal{Tectonophysics}
\bvolume{424}
\bpages{291--306}.
\bid{doi={10.1016/j.tecto.2006.03.038}}
\end{barticle}
%
\bptok{imsref}%
\endbibitem

\bibitem[\protect\citeauthoryear{Ogata}{2006b}]{OGA06N2}
%
\begin{bmisc}[auto:STB|2014/06/18|12:29:53]
\bauthor{\bsnm{Ogata},~\bfnm{Y.}\binits{Y.}}
(\byear{2006}b).
\bhowpublished{Fortran programs statistical analysis of
seismicity---Updated version, \mbox{(SASeis2006).}
\textit{Computer Science Monograph} No.~33, The Institute of
Statistical Mathematics,
Tokyo, Japan. Available at \url{http://www.ism.ac.jp/editsec/csm/index\_j.html}.}
\end{bmisc}
%
\bptok{imsref}%
\endbibitem

\bibitem[\protect\citeauthoryear{Ogata}{2007}]{OGA07}
%
\begin{barticle}[auto:STB|2014/06/18|12:29:53]
\bauthor{\bsnm{Ogata},~\bfnm{Y.}\binits{Y.}}
(\byear{2007}).
\btitle{Seismicity and geodetic anomalies in a wide preceding the
Niigata-Ken-Chuetsu earthquake of 23 October 2004, central Japan}.
\bjournal{J. Geophys. Res.}
\bvolume{112}
\bpages{B10301}.
\bid{doi={10.1029/2006JB004697}}
\end{barticle}
%
\bptok{imsref}%
\endbibitem

\bibitem[\protect\citeauthoryear{Ogata}{2010}]{OGA10}
%
\begin{barticle}[auto:STB|2014/06/18|12:29:53]
\bauthor{\bsnm{Ogata},~\bfnm{Y.}\binits{Y.}}
(\byear{2010}).
\btitle{Anomalies of seismic activity and transient crustal
deformations preceding the 2005 M7.0 earthquake west of Fukuoka}.
\bjournal{Pure Appl. Geophys.}
\bvolume{167}
\bpages{1115--1127}.
\bid{doi={10.1007/s00024-010-0096-y}}
\end{barticle}
%
\bptok{imsref}%
\endbibitem

\bibitem[\protect\citeauthoryear{Ogata}{2011a}]{OGA11N1}
%
\begin{barticle}[auto:STB|2014/06/18|12:29:53]
\bauthor{\bsnm{Ogata},~\bfnm{Y.}\binits{Y.}}
(\byear{2011}a).
\btitle{Long-term probability forecast of the regional seismicity that
was induced by the M9 Tohoku-Oki earthquake}.
\bjournal{Report of the Coordinating Committee for Earthquake Prediction}
\bvolume{88}
\bpages{92--99}.
\end{barticle}
%
\bptok{imsref}%
\endbibitem

\bibitem[\protect\citeauthoryear{Ogata}{2011b}]{OGA11N2}
%
\begin{barticle}[auto:STB|2014/06/18|12:29:53]
\bauthor{\bsnm{Ogata},~\bfnm{Y.}\binits{Y.}}
(\byear{2011}b).
\btitle{Significant improvements of the space--time ETAS model for
forecasting of accurate baseline seismicity}.
\bjournal{Earth Planets Space}
\bvolume{63}
\bpages{217--229}.
\bid{doi={10.5047/eps.2010.09.001}}
\end{barticle}
%
\bptok{imsref}%
\endbibitem

\bibitem[\protect\citeauthoryear{Ogata}{2012}]{OGA12}
%
\begin{barticle}[auto:STB|2014/06/18|12:29:53]
\bauthor{\bsnm{Ogata},~\bfnm{Y.}\binits{Y.}}
(\byear{2012}).
\btitle{Tohoku earthquake aftershock activity (in Japanese)}.
\bjournal{Report of the Coordinating Committee for Earthquake Prediction}
\bvolume{88}
\bpages{100--103}.
\end{barticle}
%
\bptok{imsref}%
\endbibitem

\bibitem[\protect\citeauthoryear{Ogata, Jones and Toda}{2003}]{OGAJONTOD03}
%
\begin{barticle}[auto:STB|2014/06/18|12:29:53]
\bauthor{\bsnm{Ogata},~\bfnm{Y.}\binits{Y.}},
\bauthor{\bsnm{Jones},~\bfnm{L.~M.}\binits{L.~M.}} \AND
\bauthor{\bsnm{Toda},~\bfnm{S.}\binits{S.}}
(\byear{2003}).
\btitle{When and where the aftershock activity was depressed:
Contrasting decay patterns of the proximate large earthquakes in
southern California}.
\bjournal{J.~Geophys. Res.}
\bvolume{108}
\bpages{B6, 2318}.
\end{barticle}
%
\bptok{imsref}%
\endbibitem

\bibitem[\protect\citeauthoryear{Ogata and Katsura}{1993}]{OGAKAT93}
%
\begin{barticle}[auto:STB|2014/06/18|12:29:53]
\bauthor{\bsnm{Ogata},~\bfnm{Y.}\binits{Y.}} \AND
\bauthor{\bsnm{Katsura},~\bfnm{K.}\binits{K.}}
(\byear{1993}).
\btitle{Analysis of temporal and special heterogeneity of magnitude
frequency distribution inferred from earthquake catalogues}.
\bjournal{Geophys. J. Int.}
\bvolume{113}
\bpages{727--738}.
\end{barticle}
%
\bptok{imsref}%
\endbibitem

\bibitem[\protect\citeauthoryear{Ogata and Katsura}{2006}]{OGAKAT06}
%
\begin{barticle}[auto:STB|2014/06/18|12:29:53]
\bauthor{\bsnm{Ogata},~\bfnm{Y.}\binits{Y.}} \AND
\bauthor{\bsnm{Katsura},~\bfnm{K.}\binits{K.}}
(\byear{2006}).
\btitle{Immediate and updated forecasting of aftershock hazard}.
\bjournal{Geophys. Res. Lett.}
\bvolume{33}
\bpages{L10305}.
\end{barticle}
%
\bptok{imsref}%
\endbibitem

\bibitem[\protect\citeauthoryear{Ogata, Katsura and
Tanemura}{2003}]{OgaKatTan03}
%
\begin{barticle}[mr]
\bauthor{\bsnm{Ogata},~\bfnm{Yosihiko}\binits{Y.}},
\bauthor{\bsnm{Katsura},~\bfnm{Koichi}\binits{K.}} \AND
\bauthor{\bsnm{Tanemura},~\bfnm{Masaharu}\binits{M.}}
(\byear{2003}).
\btitle{Modelling heterogeneous space--time occurrences of earthquakes
and its residual analysis}.
\bjournal{J. Roy. Statist. Soc. Ser. C}
\bvolume{52}
\bpages{499--509}.
\bid{doi={10.1111/1467-9876.00420}, issn={0035-9254}, mr={2012973}}
\end{barticle}
%
\bptok{imsref}%
\endbibitem

\bibitem[\protect\citeauthoryear{Okutani and Ide}{2011}]{OKUIDE11}
%
\begin{barticle}[auto:STB|2014/06/18|12:29:53]
\bauthor{\bsnm{Okutani},~\bfnm{T.}\binits{T.}} \AND
\bauthor{\bsnm{Ide},~\bfnm{S.}\binits{S.}}
(\byear{2011}).
\btitle{Statistic analysis of swarm activities around the Boso
Peninsula, Japan: Slow slip events beneath Tokyo Bay?}
\bjournal{Earth Planets Space}
\bvolume{63}
\bpages{419--426}.
\bid{doi={10.5047/eps.2011.02.010}}
\end{barticle}
%
\bptok{imsref}%
\endbibitem

\bibitem[\protect\citeauthoryear{Omi et~al.}{2013}]{OMIetal13}
%
\begin{barticle}[auto:STB|2014/06/18|12:29:53]
\bauthor{\bsnm{Omi},~\bfnm{T.}\binits{T.}},
\bauthor{\bsnm{Ogata},~\bfnm{Y.}\binits{Y.}},
\bauthor{\bsnm{Hirata},~\bfnm{Y.}\binits{Y.}} \AND
\bauthor{\bsnm{Aihara},~\bfnm{K.}\binits{K.}}
(\byear{2013}).
\btitle{Forecasting large aftershocks within one day after the main shock}.
\bjournal{Sci. Rep.}
\bvolume{3}
\bpages{2218}.
\bid{doi={10.1038/srep02218}}
\end{barticle}
%
\bptok{imsref}%
\endbibitem

\bibitem[\protect\citeauthoryear{Peng, Schoenberg and Woods}{2005}]{PenSchWoo05}
%
\begin{barticle}[mr]
\bauthor{\bsnm{Peng},~\bfnm{Roger~D.}\binits{R.~D.}},
\bauthor{\bsnm{Schoenberg},~\bfnm{Frederic~Paik}\binits{F.~P.}} \AND
\bauthor{\bsnm{Woods},~\bfnm{James~A.}\binits{J.~A.}}
(\byear{2005}).
\btitle{A space--time conditional intensity model for evaluating a
wildfire hazard index}.
\bjournal{J. Amer. Statist. Assoc.}
\bvolume{100}
\bpages{26--35}.
\bid{doi={10.1198/016214504000001763}, issn={0162-1459}, mr={2166067}}
\end{barticle}
%
\bptok{imsref}%
\endbibitem

\bibitem[\protect\citeauthoryear{Reasenberg}{1985}]{REA85}
%
\begin{barticle}[auto:STB|2014/06/18|12:29:53]
\bauthor{\bsnm{Reasenberg},~\bfnm{P.}\binits{P.}}
(\byear{1985}).
\btitle{Second-order moment of central California seismicity, 1969--1982}.
\bjournal{J. Geophys. Res.}
\bvolume{90}
\bpages{B7, 5479--5495}.
\bid{doi={10.1029/JB090iB07p05479}}
\end{barticle}
%
\bptok{imsref}%
\endbibitem

\bibitem[\protect\citeauthoryear{Schoenberg, Peng and Woods}{2003}]{SCHPENWOO03}
%
\begin{barticle}[auto:STB|2014/06/18|12:29:53]
\bauthor{\bsnm{Schoenberg},~\bfnm{F.~P.}\binits{F.~P.}},
\bauthor{\bsnm{Peng},~\bfnm{R.}\binits{R.}} \AND
\bauthor{\bsnm{Woods},~\bfnm{J.}\binits{J.}}
(\byear{2003}).
\btitle{On the distribution of wild fire sizes}.
\bjournal{Environmetrics}
\bvolume{14}
\bpages{583--592}.
\end{barticle}
%
\bptok{imsref}%
\endbibitem

\bibitem[\protect\citeauthoryear{Terakawa, Hashimoto and Matsu'ura}{2013}]{TERHASMAT13}
%
\begin{barticle}[auto:STB|2014/06/18|12:29:53]
\bauthor{\bsnm{Terakawa},~\bfnm{T.}\binits{T.}},
\bauthor{\bsnm{Hashimoto},~\bfnm{C.}\binits{C.}} \AND
\bauthor{\bsnm{Matsu'ura},~\bfnm{M.}\binits{M.}}
(\byear{2013}).
\btitle{Changes in seismic activity following the 2011 Tohoku-Oki
earthquake: Effects of pore fluid pressure}.
\bjournal{Earth Planet. Sci. Lett.}
\bvolume{365}
\bpages{17--24}.
\end{barticle}
%
\bptok{imsref}%
\endbibitem

\bibitem[\protect\citeauthoryear{Terakawa and Matsu'uara}{2010}]{TERMAT}
%
\begin{barticle}[auto:STB|2014/06/18|12:29:53]
\bauthor{\bsnm{Terakawa},~\bfnm{T.}\binits{T.}} \AND
\bauthor{\bsnm{Matsu'uara},~\bfnm{M.}\binits{M.}}
(\byear{2010}).
\btitle{The 3-d tectonic stress fields in and around Japan inverted
from centroid moment tensor data of seismic events}.
\bjournal{Tectonics}
\bvolume{29}
\bpages{(TC6008)}.
\end{barticle}
%
\bptok{imsref}%
\endbibitem

\bibitem[\protect\citeauthoryear{Terakawa, Miller and
Deichmann}{2012}]{TERMILDEI12}
%
\begin{barticle}[auto:STB|2014/06/18|12:29:53]
\bauthor{\bsnm{Terakawa},~\bfnm{T.}\binits{T.}},
\bauthor{\bsnm{Miller},~\bfnm{S.}\binits{S.}} \AND
\bauthor{\bsnm{Deichmann},~\bfnm{N.}\binits{N.}}
(\byear{2012}).
\btitle{High fluid pressure and triggered earthquakes in the enhanced
geothermal system in Basel, Switzerland}.
\bjournal{J. Geophys. Res.}
\bvolume{117}
\bpages{B07305, 15~pp.}
\end{barticle}
%
\bptok{imsref}%
\endbibitem

\bibitem[\protect\citeauthoryear{Toda, Lian and Ross}{2011}]{TODLIAROS11}
%
\begin{barticle}[auto:STB|2014/06/18|12:29:53]
\bauthor{\bsnm{Toda},~\bfnm{S.}\binits{S.}},
\bauthor{\bsnm{Lian},~\bfnm{L.}\binits{L.}} \AND
\bauthor{\bsnm{Ross},~\bfnm{S.}\binits{S.}}
(\byear{2011}).
\btitle{Using the 2011~M${}={}$9.0 Tohoku earthquake to test the
Coulomb stress triggering hypothesis and to calculate faults brought
closer to failure}.
\bjournal{Earth Planets Space}
\bvolume{63}
\bpages{725--730}.
\end{barticle}
%
\bptok{imsref}%
\endbibitem

\bibitem[\protect\citeauthoryear{Toda, Stein and Jian}{2011}]{TODSTEJIA11}
%
\begin{barticle}[auto:STB|2014/06/18|12:29:53]
\bauthor{\bsnm{Toda},~\bfnm{S.}\binits{S.}},
\bauthor{\bsnm{Stein},~\bfnm{R.~S.}\binits{R.~S.}} \AND
\bauthor{\bsnm{Jian},~\bfnm{L.}\binits{L.}}
(\byear{2011}).
\btitle{Widespread seismicity excitation throughout central Japan
following the 2011~M${}={}$9.0
Tohoku earthquake, and its interpretation in terms of Coulomb stress transfer}.
\bjournal{Geophys. Res. Lett.}
\bvolume{38}
\bpages{L00G03}.
\end{barticle}
%
\bptok{imsref}%
\endbibitem

\bibitem[\protect\citeauthoryear{Tsuruoka}{1996}]{TSU96}
%
\begin{barticle}[auto:STB|2014/06/18|12:29:53]
\bauthor{\bsnm{Tsuruoka},~\bfnm{H.}\binits{H.}}
(\byear{1996}).
\btitle{Development of seismicity analysis software on workstation (in
Japanese)}.
\bjournal{Tech. Res. Rep.}
\bvolume{2}
\bpages{34--42}.
\bnote{Earthq. Res. Inst., Univ. of Tokyo, Tokyo}.
\end{barticle}
%
\bptok{imsref}%
\endbibitem

\bibitem[\protect\citeauthoryear{Utsu}{1961}]{UTS61}
%
\begin{barticle}[auto:STB|2014/06/18|12:29:53]
\bauthor{\bsnm{Utsu},~\bfnm{T.}\binits{T.}}
(\byear{1961}).
\btitle{Statistical study on the occurrence of aftershocks}.
\bjournal{Geophys. Mag.}
\bvolume{30}
\bpages{521--605}.
\end{barticle}
%
\bptok{imsref}%
\endbibitem

\bibitem[\protect\citeauthoryear{Utsu}{1962}]{UTS62}
%
\begin{barticle}[auto:STB|2014/06/18|12:29:53]
\bauthor{\bsnm{Utsu},~\bfnm{T.}\binits{T.}}
(\byear{1962}).
\btitle{On the nature of three Alaskan aftershock sequences of 1957 and 1958}.
\bjournal{Bull. Seismol. Soc. Amer.}
\bvolume{52}
\bpages{279--297}.
\end{barticle}
%
\bptok{imsref}%
\endbibitem

\bibitem[\protect\citeauthoryear{Utsu}{1969}]{UTS69}
%
\begin{barticle}[auto:STB|2014/06/18|12:29:53]
\bauthor{\bsnm{Utsu},~\bfnm{T.}\binits{T.}}
(\byear{1969}).
\btitle{Aftershocks and earthquake statistics (I)---Some parameters
which characterize an aftershock sequence and their interrelations}.
\bjournal{J. Fac. Sci. Hokkaido Univ., Ser. VII}
\bvolume{3}
\bpages{129--195}.
\end{barticle}
%
\bptok{imsref}%
\endbibitem

\bibitem[\protect\citeauthoryear{Utsu}{1970}]{UTS70}
%
\begin{barticle}[auto:STB|2014/06/18|12:29:53]
\bauthor{\bsnm{Utsu},~\bfnm{T.}\binits{T.}}
(\byear{1970}).
\btitle{Aftershocks and earthquake statistics (II)---Further
investigation of aftershocks and other earthquake sequences based on a
new classification of earthquake sequences}.
\bjournal{J. Fac. Sci. Hokkaido Univ., Ser. VII}
\bvolume{3}
\bpages{197--266}.
\end{barticle}
%
\bptok{imsref}%
\endbibitem

\bibitem[\protect\citeauthoryear{Utsu}{1971}]{UTS71}
%
\begin{barticle}[auto:STB|2014/06/18|12:29:53]
\bauthor{\bsnm{Utsu},~\bfnm{T.}\binits{T.}}
(\byear{1971}).
\btitle{Aftershocks and earthquake statistics (III)---Analyses of the
distribution of earthquakes in magnitude,
time, and space with special consideration to clustering
characteristics of earthquake occurrence (1)}.
\bjournal{J. Fac. Sci. Hokkaido Univ., Ser. VII}
\bvolume{3}
\bpages{379--441}.
\end{barticle}
%
\bptok{imsref}%
\endbibitem

\bibitem[\protect\citeauthoryear{Utsu}{1972}]{UTS72}
%
\begin{barticle}[auto:STB|2014/06/18|12:29:53]
\bauthor{\bsnm{Utsu},~\bfnm{T.}\binits{T.}}
(\byear{1972}).
\btitle{Aftershocks and earthquake statistics (IV)---Analyses of the
distribution of earthquakes in magnitude,
time, and space with special consideration to clustering
characteristics of earthquake occurrence (2)}.
\bjournal{J. Fac. Sci. Hokkaido Univ., Ser. VII}
\bvolume{4}
\bpages{1--42}.
\end{barticle}
%
\bptok{imsref}%
\endbibitem

\bibitem[\protect\citeauthoryear{Utsu, Ogata and Matsu'ura}{1995}]{UTSOGAMAT95}
%
\begin{barticle}[auto:STB|2014/06/18|12:29:53]
\bauthor{\bsnm{Utsu},~\bfnm{T.}\binits{T.}},
\bauthor{\bsnm{Ogata},~\bfnm{Y.}\binits{Y.}} \AND
\bauthor{\bsnm{Matsu'ura},~\bfnm{R.~S.}\binits{R.~S.}}
(\byear{1995}).
\btitle{The centenary of the Omori formula for a decay law of
aftershock activity}.
\bjournal{J. Seismol. Soc. Japan}
\bvolume{7}
\bpages{233--240}.
\end{barticle}
%
\bptok{imsref}%
\endbibitem

\bibitem[\protect\citeauthoryear{Utsu and Seki}{1955}]{UTSSEK55}
%
\begin{barticle}[auto:STB|2014/06/18|12:29:53]
\bauthor{\bsnm{Utsu},~\bfnm{T.}\binits{T.}} \AND
\bauthor{\bsnm{Seki},~\bfnm{A.}\binits{A.}}
(\byear{1955}).
\btitle{A relation between the area of after-shock region and the
energy of main shock (in Japanese)}.
\bjournal{Zisin (2)}
\bvolume{7}
\bpages{233--240}.
\end{barticle}
%
\bptok{imsref}%
\endbibitem

\bibitem[\protect\citeauthoryear{Vere-Jones}{1970}]{Ver70}
%
\begin{barticle}[mr]
\bauthor{\bsnm{Vere-Jones},~\bfnm{D.}\binits{D.}}
(\byear{1970}).
\btitle{Stochastic models for earthquake occurrence}.
\bjournal{J. Roy. Statist. Soc. Ser. B}
\bvolume{32}
\bpages{1--62}.
\bid{issn={0035-9246}, mr={0272087}}
\bptnote{check related}%
\end{barticle}
%
\bptok{imsref}%
\endbibitem

\bibitem[\protect\citeauthoryear{Vere-Jones and Davies}{1966}]{VERDAV66}
%
\begin{barticle}[auto:STB|2014/06/18|12:29:53]
\bauthor{\bsnm{Vere-Jones},~\bfnm{D.}\binits{D.}} \AND
\bauthor{\bsnm{Davies},~\bfnm{R.~B.}\binits{R.~B.}}
(\byear{1966}).
\btitle{A statistical study of earthquakes in the main seismic area of
New Zealand. Part II: Time series analyses}.
\bjournal{N.~Z. J.~Geol. Geophys.}
\bvolume{9}
\bpages{251--284}.
\end{barticle}
%
\bptok{imsref}%
\endbibitem

\bibitem[\protect\citeauthoryear{Zhuang and Ogata}{2006}]{ZHUOGA06}
%
\begin{barticle}[auto:STB|2014/06/18|12:29:53]
\bauthor{\bsnm{Zhuang},~\bfnm{J.}\binits{J.}} \AND
\bauthor{\bsnm{Ogata},~\bfnm{Y.}\binits{Y.}}
(\byear{2006}).
\btitle{Properties of the probability distribution associated with the
largest earthquake in a cluster and their implications to foreshocks}.
\bjournal{Phys. Rev. E}
\bvolume{73}
\bpages{046134}.
\bid{doi={10.1103/PhysRevE.73.046134}}
\end{barticle}
%
\bptok{imsref}%
\endbibitem

\bibitem[\protect\citeauthoryear{Zhuang, Ogata and Vere-Jones}{2002}]{ZhuOgaVer02}
%
\begin{barticle}[mr]
\bauthor{\bsnm{Zhuang},~\bfnm{Jiancang}\binits{J.}},
\bauthor{\bsnm{Ogata},~\bfnm{Yosihiko}\binits{Y.}} \AND
\bauthor{\bsnm{Vere-Jones},~\bfnm{David}\binits{D.}}
(\byear{2002}).
\btitle{Stochastic declustering of space--time earthquake occurrences}.
\bjournal{J. Amer. Statist. Assoc.}
\bvolume{97}
\bpages{369--380}.
\bid{doi={10.1198/016214502760046925}, issn={0162-1459}, mr={1941459}}
\end{barticle}
%
\bptok{imsref}%
\endbibitem

\bibitem[\protect\citeauthoryear{Zhuang, Ogata and
Vere-Jones}{2004}]{ZHUOGAVER04}
%
\begin{barticle}[auto:STB|2014/06/18|12:29:53]
\bauthor{\bsnm{Zhuang},~\bfnm{J.}\binits{J.}},
\bauthor{\bsnm{Ogata},~\bfnm{Y.}\binits{Y.}} \AND
\bauthor{\bsnm{Vere-Jones},~\bfnm{D.}\binits{D.}}
(\byear{2004}).
\btitle{Analyzing earthquake clustering features by using stochastic
reconstruction}.
\bjournal{J. Geophys. Res.}
\bvolume{109}
\bpages{B5, B05301}.
\end{barticle}
%
\bptok{imsref}%
\endbibitem

\end{thebibliography}
\end{document}